\documentclass[fleqn,usenatbib]{mnras}

\usepackage{newtxtext,newtxmath}
\usepackage{subfig}
\usepackage{float}
\usepackage{url}
\usepackage{multirow}
\usepackage[T1]{fontenc}

\DeclareRobustCommand{\VAN}[3]{#2}
\let\VANthebibliography\thebibliography
\def\thebibliography{\DeclareRobustCommand{\VAN}[3]{##3}\VANthebibliography}

\newcommand{\solaris}{\textit{Solaris}}
\newcommand{\ccdproc}{\textsc{ccdproc}}
\newcommand{\bk}{BK\,Ind}
\newcommand{\vara}{V889\,Ara}
\newcommand{\su}{SU\,Ind}
\newcommand{\cpd}{CPD-52\,10541}
\newcommand{\hd}{HD\,60637}
\newcommand{\tyc}{TYC\,8504-1018-1}
\newcommand{\gsc}{GSC\,08814-01026}

\usepackage{graphicx}	% Including figure files
\usepackage{amsmath}	% Advanced maths commands
\title[\solaris{} Photometric Survey]{ \solaris{} photometric survey: Search for circumbinary companions using eclipse timing variations}

\author[A. Moharana et al.]{
A. Moharana,$^{1}$\thanks{E-mail: ayushm@ncac.torun.pl}
K.~G. He{\l}miniak,$^{1}$
F. Marcadon,$^{2}$
T. Pawar,$^{1}$
G. Pawar,$^{1}$
P. Garczy\'nski,$^{3}$
J. Per{\l}a,$^{4}$
\and
S.~K. Koz{\l}owski$^{5}$, P. Sybilski$^{6}$, M. Ratajczak$^{7}$,
and M. Konacki$^{8}$
\\
$^{1}$Nicolaus Copernicus Astronomical Center, Polish Academy of Sciences, ul. Rabia\'{n}ska 8, 87-100 Toru\'{n}, Poland\\
$^{2}$Villanova University, Dept.\ of Astrophysics and Planetary Sciences, 800 East Lancaster Avenue, Villanova, PA 19085, USA\\
$^{3}$Faculty of Physics, University of Warsaw, Pasteura 5, 02-093 Warszawa, Poland \\
$^{4}$Astronomical Observatory Institute, Faculty of Physics, Adam Mickiewicz University, S{\l}oneczna 36, 60-286 Pozna\'{n}, Poland \\
$^{5}$Cilium Engineering Sp. z o.o., ul. {\L}okietka 5, 87-100 Toru\'{n}, Poland\\
$^{6}$Sybilla Technologies Sp. z o.o., ul. Toru\'{n}ska 59, 85-023 Bydgoszcz, Poland\\
$^{7}$ Astronomical Observatory, University of Warsaw, Al. Ujazdowskie 4, 00-478 Warszawa, Poland\\
$^{8}$Nicolaus Copernicus Astronomical Center, Polish Academy of Sciences, ul. Bartycka 18, 00-716 Warszawa, Poland\\
}

\date{Accepted XXX. Received YYY; in original form ZZZ}

\pubyear{2015}

\begin{document}
\label{firstpage}
\pagerange{\pageref{firstpage}--\pageref{lastpage}}
\maketitle

\begin{abstract}
Eclipse timing variations (ETV) have been a successful tool for detecting circumbinary companions to eclipsing binaries (EB). While \textit{TESS} and \textit{Kepler} have been prolific for ETV searches, they sometimes can be limited by time and sky coverage which can be addressed by specialised ground-based ETV surveys. We present the initial results from the \solaris{} photometric survey which uses four 0.5m robotic telescopes in the southern hemisphere to look for circumbinary companions. We present the method of light curve extraction, detrending, and EB modelling using the observations from the \solaris{} network. Using these light curves we extract precise eclipse timing for 7 EB and look for companions using a Lomb-Scargle periodogram search. We find two possible periodic signals for the target \gsc{}. With the system having strong activity, we check for the feasibility of orbital solutions at these two periods. We find that the $245 \pm 1$ d period is due to an M-dwarf mass companion. This makes \gsc{} a candidate compact hierarchical triple system. The other periodic signal at $146 \pm 1$ d is an artefact of stellar activity.
 \end{abstract}


%%%%%%%%%%%%%%%%%%%%%%%%%%%%%%%%%%%%%%%%%%%%%%%%%%

%%%%%%%%%%%%%%%%% BODY OF PAPER %%%%%%%%%%%%%%%%%%

\section{Introduction}

The detection of companions to stars has been crucial for understanding the formation, evolution, and dynamics of stars. It paved the way for the study of stellar multiplicity. The theory of the formation of low-mass objects like brown dwarfs and planets was also greatly influenced by observations of stellar multiplicity. 

Most of the early surveys of multiplicity began with astrometric, radial velocity, and visual detections of companions \citep{visualsrch, battenbook,rvsearch1976}. Meanwhile, \cite{fekel1981} looked at such a survey and identified an eclipsing binary (EB) in one of the stars in a resolved binary. This led to the detection of one of the first tertiary around an EB. Since then, EBs have been used extensively for multiplicity studies using photometry, in addition to other avenues of detection. Photometry itself can be used in three different ways to detect a companion to an EB: (i) third light in light curve solutions, (ii) eclipse depth variations (EDV), and (iii) eclipse timing variations (ETV). While the first method is a qualitative approach, EDV is prominent in highly dynamic systems. Meanwhile, ETV has proven to be quite reliable in observing the light travel time effect (LTTE) since the work of \cite{chandleralgol}, who first gave a possible reason for period changes in Algol.  

LTTE is observed due to the movement of the EB with respect to the barycenter of the multiple-system. It causes the eclipses to occur earlier or later than the time expected for an isolated EB.  But the variations in the eclipse timings can also be triggered by various other physical processes, such as the transfer of magnetic and orbital momentum \citep{applegate1992}, loss of angular momentum \citep{bradstreetangmom1994} and starspots \citep{kalimeris2002,balaji2015}. However, these effects have different timescales and can be distinguished from long-term monitoring of the eclipse timings (ET). This is where large-scale photometric surveys have been helpful for companion detection.  

Photometric surveys such as Optical Gravitational Lensing Experiment (OGLE; \citealt{ogle1}), All-Sky Automated Survey (ASAS; \citealt{asas}), Super Wide Angle Search for Planets (SWASP; \citealt{Swasp}),  Hungarian Automated Telescope (HAT; \citealt{hat}), and North Sky Variability Survey (NSVS; \citealt{nsvs}) have generated a large number of observations for companion detection through ETV \citep{stantiming,ogletripleshajdu}. Meanwhile, space-based photometric missions like Convection, Rotation and planetary Transits (CoRoT; \citealt{corot}),  \textit{Kepler} \citep{kepler} and Transiting Exoplanet Survey Satellite (TESS; \citealt{tess}) have been prolific in providing precise ET and circumbinary companions \citep{keplertriplesborko,corottriples,corottesseb}. 

Meanwhile, in 2010, the \solaris{} project, a specialised ground-based photometric survey, was started to target short-period EBs for ET. In this paper, we present the first results from the ETV search using this telescope network. We discuss the setup of the telescopes and the photometric pipeline in sections \ref{sec:setup} and \ref{sec:pipeline}, respectively. In sections \ref{sec:lc} and \ref{sec:etv} we describe how we use the light curves (LC) for EB modelling and ET extraction. Finally, in section \ref{sec:results} we present our results and a positive detection.

\section{Solaris Network of Telescopes}
\label{sec:setup}
Numerical simulations have shown that to detect circumbinary companions,  multiple 0.5-m telescopes are required to continuously monitor short-period binaries (binary period < 3 days) at high-cadence \citep{piotrtimingNS}. This was the idea behind the establishment of the \solaris{} network. 

The \solaris{} network is a global network of telescopes consisting of four fully autonomous observatories located in the Republic of South Africa (Solaris-1 and -2), Australia (Solaris-3), and Argentina (Solaris-4). The headquarters and main database are located in Poland. Each observatory consists of a telescope with a  0.5 m diameter primary mirror, installed on a modified German equatorial mount from Astrosysteme Austria and equipped with fast and precise direct drives and high-resolution rotary-pulsing encoders. The Solaris-3 telescope is a Schmidt-Cassegrain f/9 optical system with a corrector, while the other telescopes are Ritchey-Cretien f/15 optical systems.
For imaging, we use professional grade Andor Icon-L CCD cameras with a resolution of 2048 x 2048 pixels, thermoelectrically cooled to -70 ° C. The filter wheels allow multi-colour photometry in ten bands: U, B, V, R, I (Johnson), and u’, g’, y’ and z’ (Sloan). The commissioning of the network, its hardware, software and processing capabilities are described in \citep{Kozlowski_2017}.

\section{The photometric pipeline}
\label{sec:pipeline}
The \solaris{} observations are stored in the databases at Toru\'n, Poland. The \solaris{} pipeline processes these observations and produces multiband light curves for each target. The steps and processes that the pipeline executes are described in the subsections below.
\subsection{Inventory}
The \solaris{} survey observed around 200 different eclipsing binaries over 5 years. The initial target sample was filtered from the ASAS Catalogue based on the maximum timing precision obtainable and other filters like duration, depth of the eclipses and brightness of the targets \citep{piotrtimingNS}. To systematically look for targets with sufficient observations, we created an inventory of the observations.

 The inventory consisted of observations done with \solaris{} from June 2015 to September 2022. The targets were identified by the right ascension and declination in their image file headers and were sorted according to the number of frames in all filters and then the number of nights of observations. The inventory also contained frame lists for every target observed. These frame lists contained the full path to the observations and therefore made it easier for the photometric routine to search for the image files. 

For this paper, we filtered targets with at least 16,000 frames of observations, spread across a minimum of 30 nights. We further narrowed down this sample to 7 targets based on the quantity and quality of the eclipses in the final LC. The periods of these targets are spread from 0.7 to 2 d with a bias towards 1d period which may be due to eclipse visibility in the basic telescope scheduling but since these make up a small sample, it is tough to draw any conclusions.

\subsection{Calibration, reduction and astrometry}
 We process two types of calibration frames: (i) bias and (ii) flats. It is possible to acquire dark frames with the \solaris{} telescopes but since the CCDs are cooled while taking images, the calibration is unaffected by the lack of it. The inventory eases the process of collecting calibration frames specific to a star.
We use \ccdproc \footnote{https://github.com/astropy/ccdproc} to median-combine and create the master-bias frames. 

The flats are sky-flats and are of two distinct configurations depending on the orientation of the CCD. To create a master-flat we first select the frames with mean-counts more than 15000. The flats were also median-combined using \textsc{ccdproc}, after being bias corrected with the master-bias for the respective night. 
Though \solaris{} takes calibration frames for every night of observation, there are a few nights when it fails to do so. We added a calibration-frame matching module to improve the yield of image frames. For every science frame, we calculate the nearest possible night (past and future) from the available calibration frames.  We reject the nights which do not have the best match within a 50-day time difference.  For the sake of consistency, we rotate the frames to a single orientation using the rotation module available in \textsc{numpy\footnote{https://numpy.org/doc/stable/reference/generated/numpy.rot90.html}}. After this we use the \ccdproc{} module \texttt{gain\_correct}, before proceeding to the standard calibration process. This gives us the final science image. 
Before proceeding to the photometry of the science frames, we apply astrometric corrections. This is to facilitate the selection of the correct target and reference star in all the photometric frames. We implement the blind astrometry algorithm \citep{astrometry} from \textit{Astrometry.net}. The frames where the algorithm fails to identify the stars are rejected in further steps. 
%====================================================

\subsection{Photometry}
Since the observations are spread over a large period,  some of the observations are affected by seeing. Having a single aperture for different nights does not work in this case and neither does creating a single PSF  (due to the lack of stars in the frames to generate a reliable PSF). Therefore, we use a variable aperture which is calculated by evaluating a PSF for each frame using three stars (one target and two reference stars). For this, we define an initial box around the stars (using the astrometric coordinates). The size of the box is estimated by checking the distance between the target and any nearby star to avoid any contamination in the box. We then fit a 2D Gaussian profile to obtain the X and Y centroids, semi-major ($a$) and semi-minor axes ($b$), and the orientation of this profile. We then centre our aperture on the centroids and the radius for the target aperture ($r_{T}$) is calculated as,
\begin{equation}
    r_{T} = 1.6 \times \sqrt{\mathrm{FWHM}_a^2+\mathrm{FWHM}_b^2}
\end{equation}
where $\mathrm{FWHM}_a = 2.35\times a$ and $\mathrm{FWHM}_b = 2.35\times b$. The background was calculated using an annular aperture with the inner radius ($r_\mathrm{Bin}$) and the outer radius ($r_\mathrm{Bout}$) given as, 

\begin{equation}
\begin{split}
    r_\mathrm{Bin}&=r_T +2  \, ; \\
   r_\mathrm{Bout}&=\sqrt{3\times r_T^2+r_\mathrm{Bin}^2}
\end{split}
\end{equation}

The photometric errors were initially calculated from the routines in \textsc{photutils} \citep{photutils} with the target flux and background flux as inputs. However, these errors were overestimated. Therefore, we calculated an overestimation scaling for different targets using the Markov Chain Monte Carlo (MCMC) method (see Sec.\ref{sec:etv}).

The pipeline produces the final LC with barycentric corrections based on \cite{eastman2014} and was implemented using \textsc{barycorrpy}\footnote{https://pypi.org/project/barycorrpy/}. This was incorporated by calculating the corrections for all 4 telescopes.

\section{Light curves}
\label{sec:lc}
\subsection{Detrending}
The pipeline generates LC spanning over 5 years of observations using relative photometry from 4 telescopes. This, therefore, resulted in various instrumental trends across the timescale of observations. We used \textsc{wotan} \citep{wotan} in two steps to get out the final LC. First, we clear out long-term trends with a window with a size 60 times the EB period. Then we clean the short-term trends using a window sized between 1/6 to 1/2 of the EB period, depending on the scale of variations in each target. We used two methods in our detrending: \texttt{biweight} \citep{biweight} and \texttt{ramsay} \citep{ramsay}. Their use in our two steps varied for different iterations. We finally selected the output LC with the least variations and also found that the best combination of a window and a particular method varied for different targets.  

\subsection{Eclipsing Binary Modelling}
\begin{figure*}
\begin{tabular}{cc}
  \includegraphics[width=0.4\textwidth]{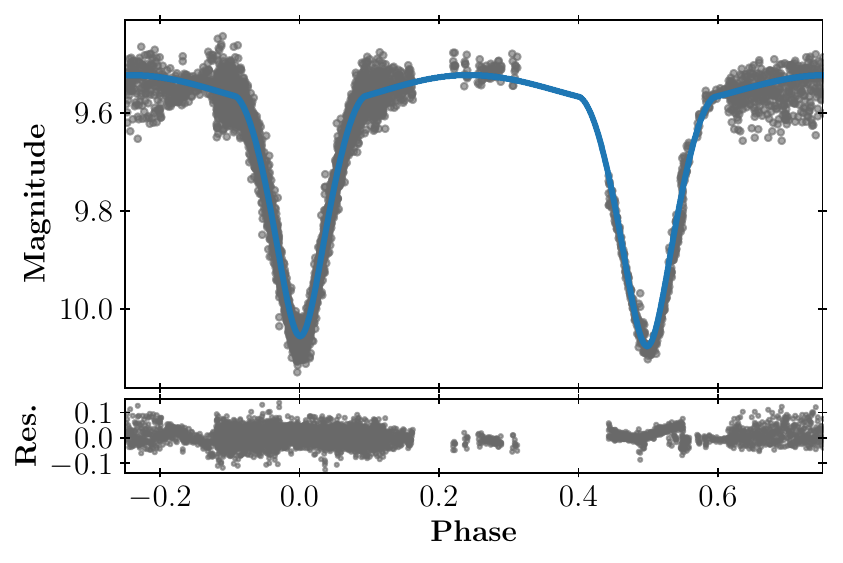} &   \includegraphics[width=0.4\textwidth]{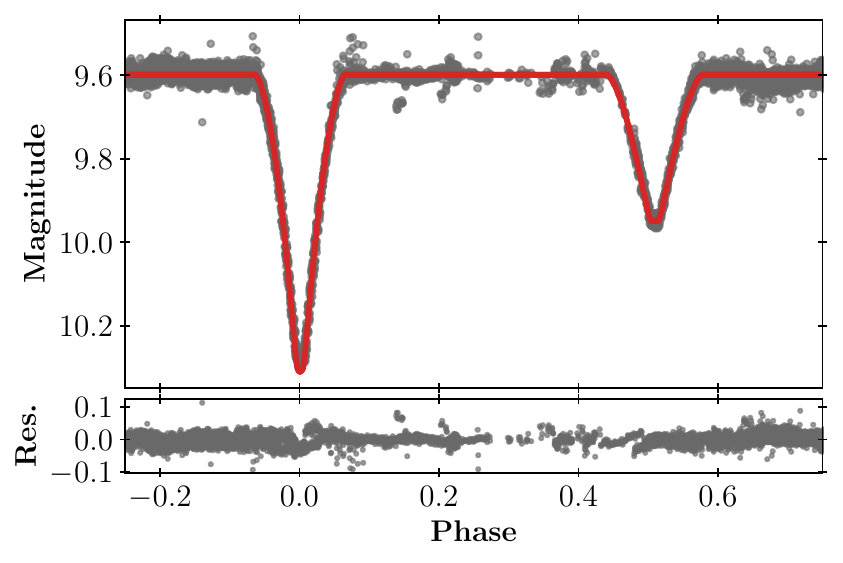} \\
(a) SU Ind:  V-band, $P = 0.9863572$  d & (b) CPD-52 10541:  I-band, $P = 1.0160554$ d\\[6pt]
 \includegraphics[width=0.4\textwidth]{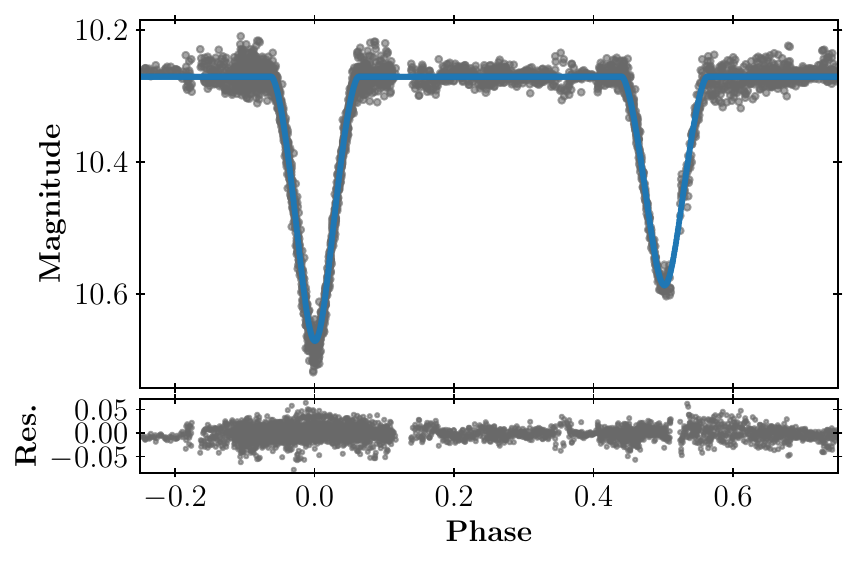} &   \includegraphics[width=0.4\textwidth]{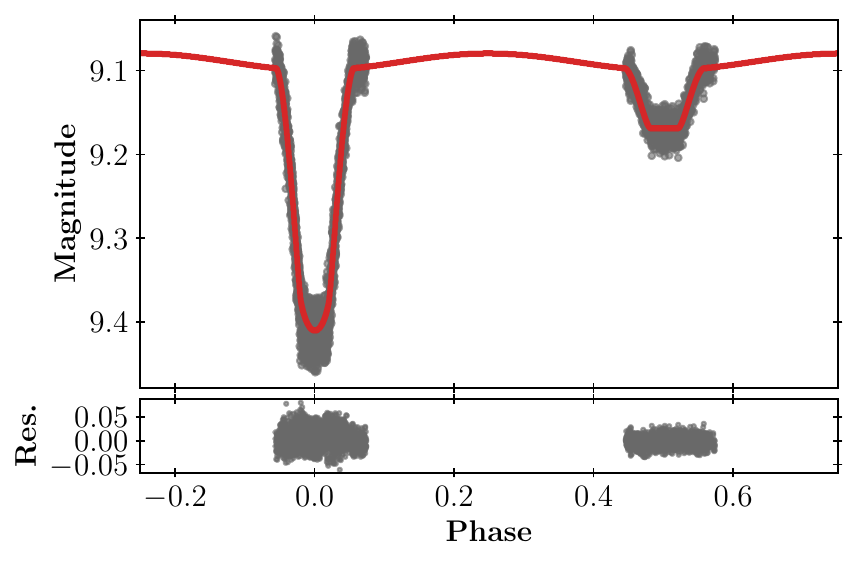} \\
(c) BK Ind:  V-band, $P = 1.112484$ d& (d) HD 60637:  I-band, $P = 1.446247$ d\\[6pt]
 \includegraphics[width=0.4\textwidth]{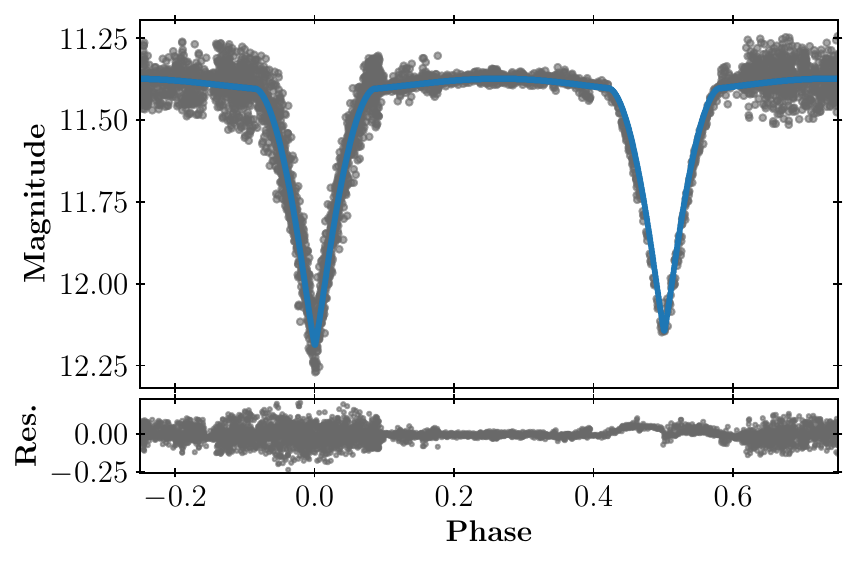} &   \includegraphics[width=0.4\textwidth]{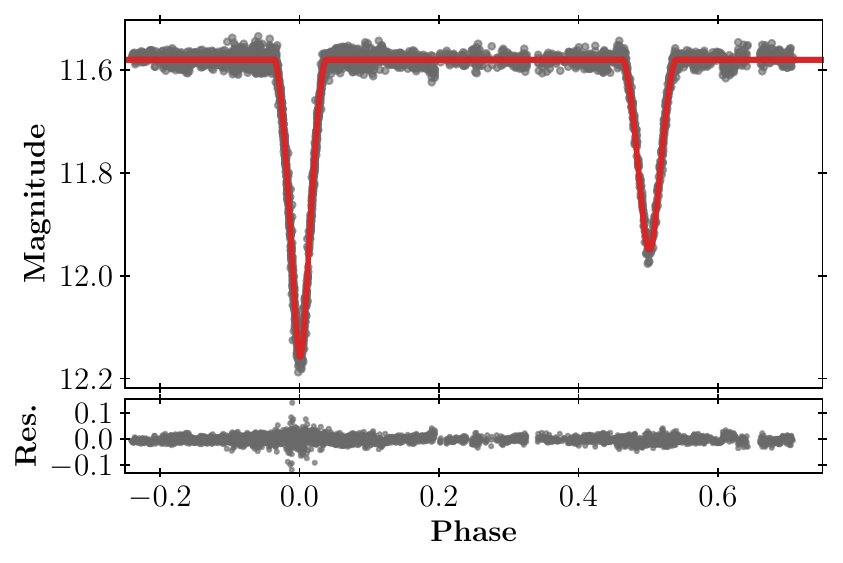} \\
(e)  V889 Ara:  V-band, $P = 1.053322$ d & (f)  TYC 8504-1018-1:  I-band, $P = 1.933535$ d \\[6pt]
\multicolumn{2}{c}{\includegraphics[width=0.4\textwidth]{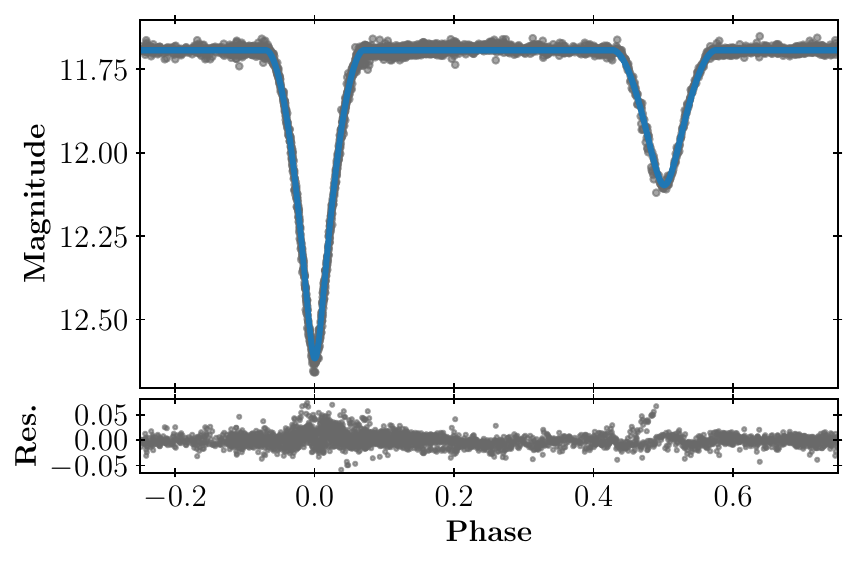}} \\
\multicolumn{2}{c}{(g)  \gsc{}:  V-band, $P = 0.702444$ d} \\[6pt]
\end{tabular}
\caption{\textsc{jktebop} LC models for the detrended \textit{Solaris} LC. The fits are on the LC segments with errors $<2\%$. Red and blue lines denote models for I and V filters respectively.}
\label{fig:lcfits}
\end{figure*}

We used version 40 of \textsc{jktebop} \citep{jktebop} for the modelling of the detrended LC. We select the LC points with minimal errors for LC modelling. We started our modelling with initial estimates of the time of periastron passage ($T_{0}$) close to the first visible primary eclipse and a period ($P$) close to initial periodogram estimates. For some targets, we used some initial parameters from RV modelling (see Appendix \ref{appx:RV}). We also used initial parameters from the literature if available. We kept the $P$, scale-factor ($S$) which controls the level of the out-of-eclipse fluxes in the LC, secondary to primary surface brightness ratio ($J$), inclination ($i$), radius-ratio ($k=r_{2}/r_{1}$), and sum of fractional radii ($r_{1}+r_{2}$) free for the first round of optimisation. After this initial optimisation, we kept the $S$ fixed and in addition to the previous set of free parameters,  $e\cos{\omega}$ and $e\sin{\omega}$ (where $e$ is eccentricity and $\omega$ is the argument of periastron) were made free. 
Logarithmic limb-darkening (LD) was assumed and approximate values for the coefficients were estimated from \cite{claret} using prior information, if available. For the rest, we used estimates for a solar-like star. Few iterations with LD coefficients free were executed before fixing them to a robust value obtained from these iterations. 
Again multiple iterations were made with the above-mentioned parameters free, before proceeding to error estimation using the Monte Carlo (MC) method in \textsc{jktebop}. For this setup, we also kept the third light parameter ($l_3$) free and then ran the MC module in \textsc{jktebop} for 10,000 iterations. The best-fit LC models for all the targets are shown in Fig.\ref{fig:lcfits}. The orbital and stellar parameter estimates are given in Tables \ref{tab:orbitparams} and \ref{tab:stellarparams}.

\section{Eclipse timing variations}
\label{sec:etv}
A tertiary around a binary system can produce three different classes of perturbation \citep{browndynamics}. They are (a) short-period perturbations in order of inner orbital period ($P_\mathrm{1}$), (b) long-period perturbations in order of outer orbital period ($P_\mathrm{2}$), and (c) apse-node perturbations of order ($P^2_\mathrm{2}/P_\mathrm{1}$).
The current setup of \solaris{} cannot constrain the short-period perturbations but long-term monitoring of the EB allows us to look for long-period perturbations. 

\subsection{Minima time extraction}

\begin{figure}
    \centering
    \includegraphics[width=\columnwidth]{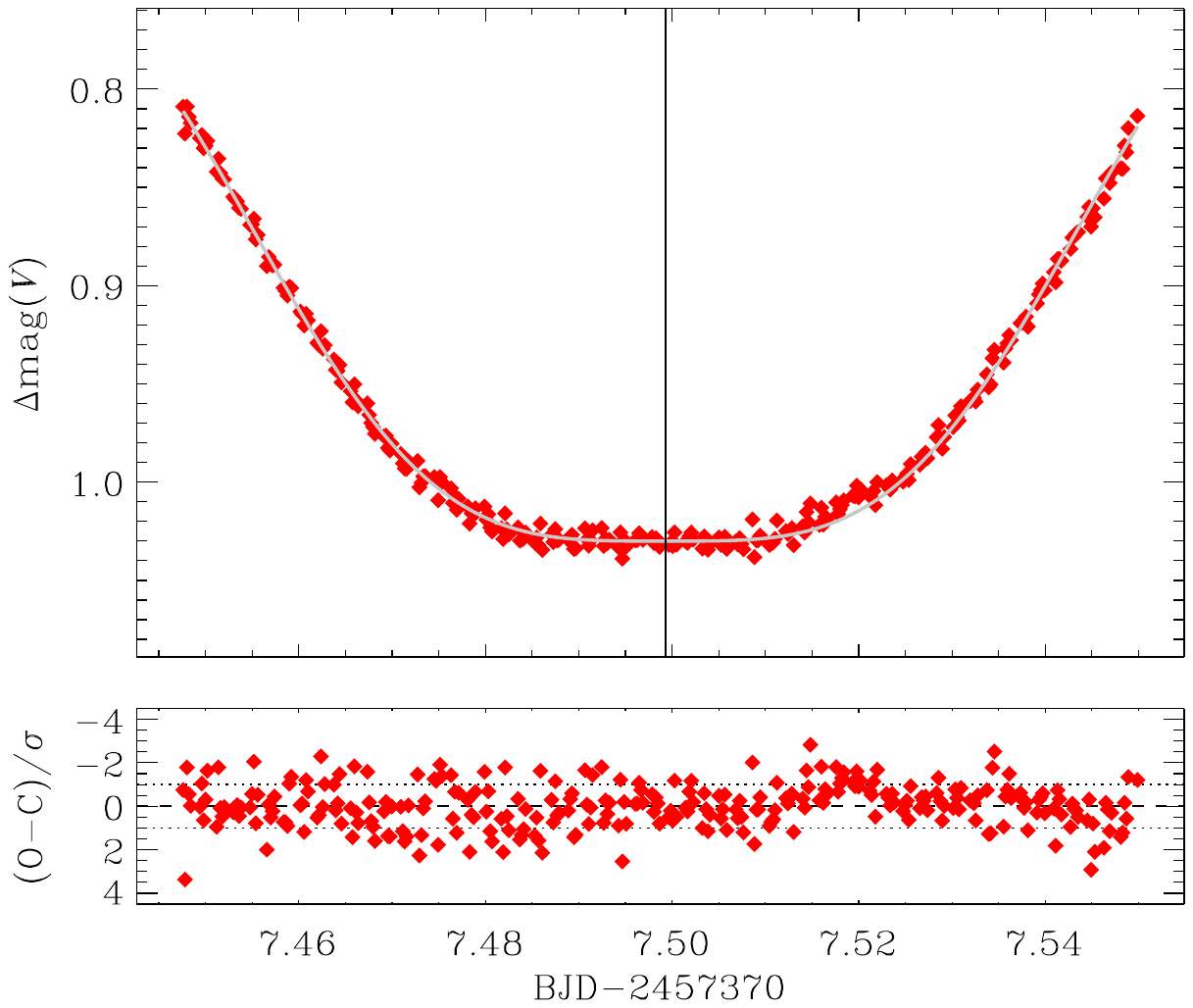}
    \caption{Eclipse fitting of the \solaris{} observations of \hd{} (primary eclipse) using our eclipse fitting code.}
    \label{fig:eclfithd}
\end{figure}

To obtain the eclipse times we follow the formalism from \cite{mikueclmodel} to fit the morphology of individual eclipses with the following function:
\begin{equation}
    f(T^j_i,\theta) = \alpha_0 + \alpha_1 \, \psi(T^j_i,T^j_0,d,\Gamma),
    \label{eq:ecl_model_1}
\end{equation}
where  $\Gamma$ defines the kurtosis of eclipse and $d$ defines the depth. $\alpha_0$ is the  magnitude zero-point shift and $\alpha_1$ is a multiplicative constant of the eclipse profile function ($\psi$), which is written as:
\begin{equation}
    \psi(T^j_i,T^j_0,d,\Gamma) = 1-\bigg\{ 1-\exp \bigg[ 1-\cosh \bigg( \frac{T^j_i-T^j_0}{d} \bigg) \bigg] \bigg\}^\Gamma.
    \label{eq:ecl_model_2}
\end{equation}
where the time of minimum of the j-component ($T^j_0$; where j=p for primary and j=s for secondary) is the parameter that is important for this work. 
This is incorporated in an eclipse fitting code used in \cite{fredv}, which can effectively calculate the morphology of eclipses and extract ET (Fig.\ref{fig:eclfithd}). The errors in the measurements are calculated using MCMC fitting. This fitting also corrects for the overestimation of magnitude errors of the LC. This is done for one eclipse and then the calculated correction factor is applied to all of the photometric observations of a target.

We also checked for timing error estimates (TEE) for our targets using the basic form of the TEE as given in \cite{tee},
\begin{equation}
    \sigma_t =\frac{\sigma_{F_{\nabla}} T_{\nabla}}{2\Delta F}
\end{equation}

where, $\sigma_{F_{\nabla}}$ is the photometric error on the timescale of ingress and egress (varying from 0.005 to 0.4 in flux units), $T_{\nabla}$ is the summed duration of ingress and egress (assumed to be equal to eclipse duration) and $\Delta F$ is the depth of the eclipse in flux units. We get TEE in the range of 0.1-1.8 times the MCMC errors, varying mostly due to the $\sigma_{F_{\nabla}}$ (which is dependent on the out-of-eclipse noise). We adopt only the MCMC errors as they are more cautious estimates for most of our ET.
The ET were obtained in two filters (I and V) for both primary and secondary eclipses. 
\subsection{Calculating variations}
\begin{figure*}
\begin{tabular}{cc}
  \includegraphics[width=0.45\textwidth]{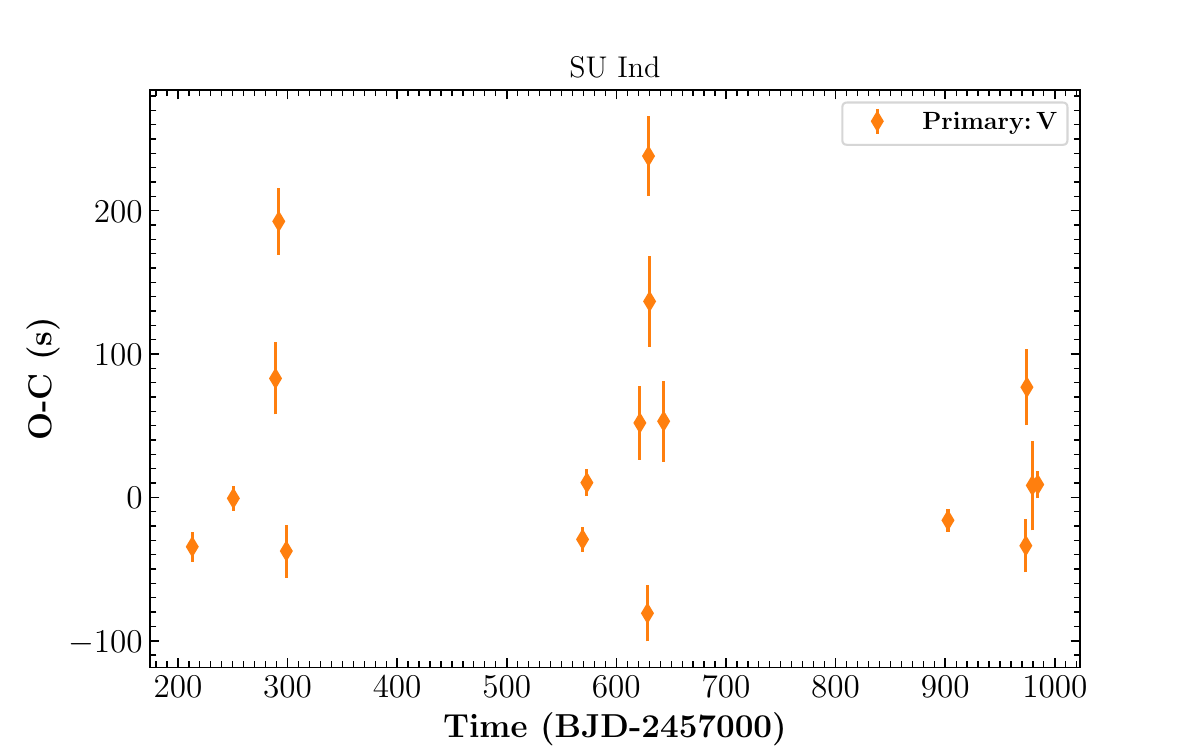} &   \includegraphics[width=0.45\textwidth]{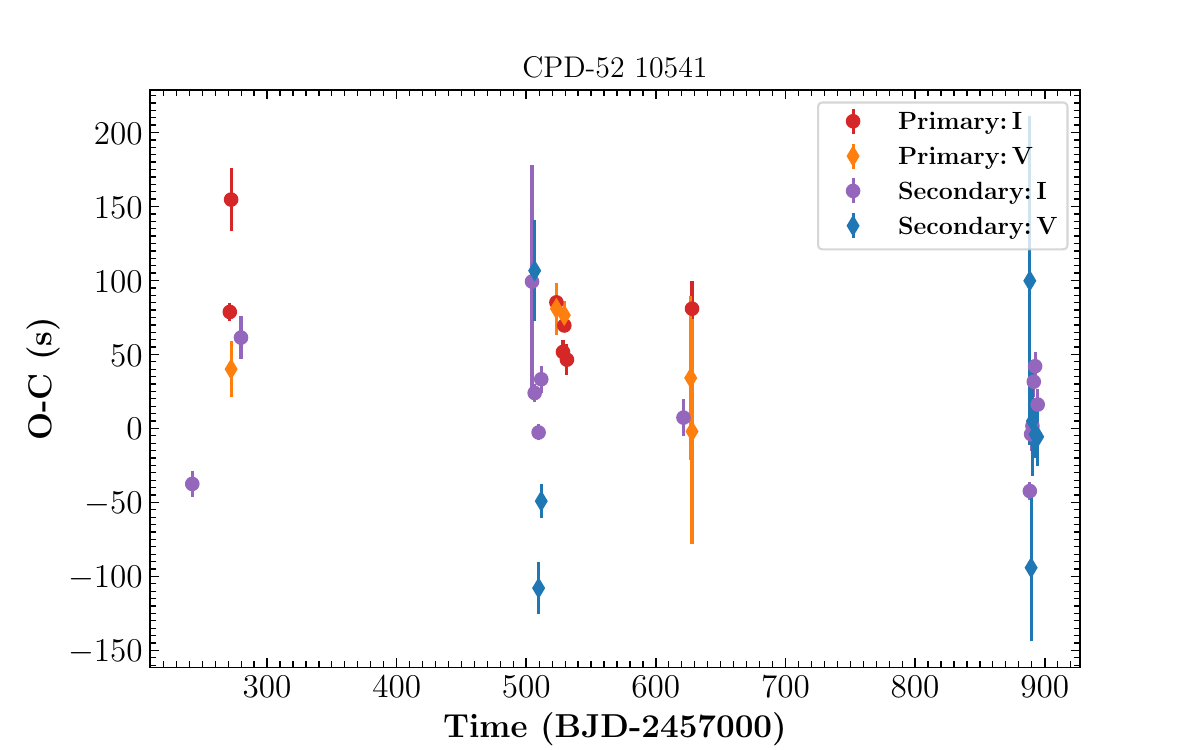} \\
(a) SU Ind:  $P =0.9863581 $  d & (b) CPD-52 10541:   $P = 1.0160550$ d\\[6pt]
 \includegraphics[width=0.45\textwidth]{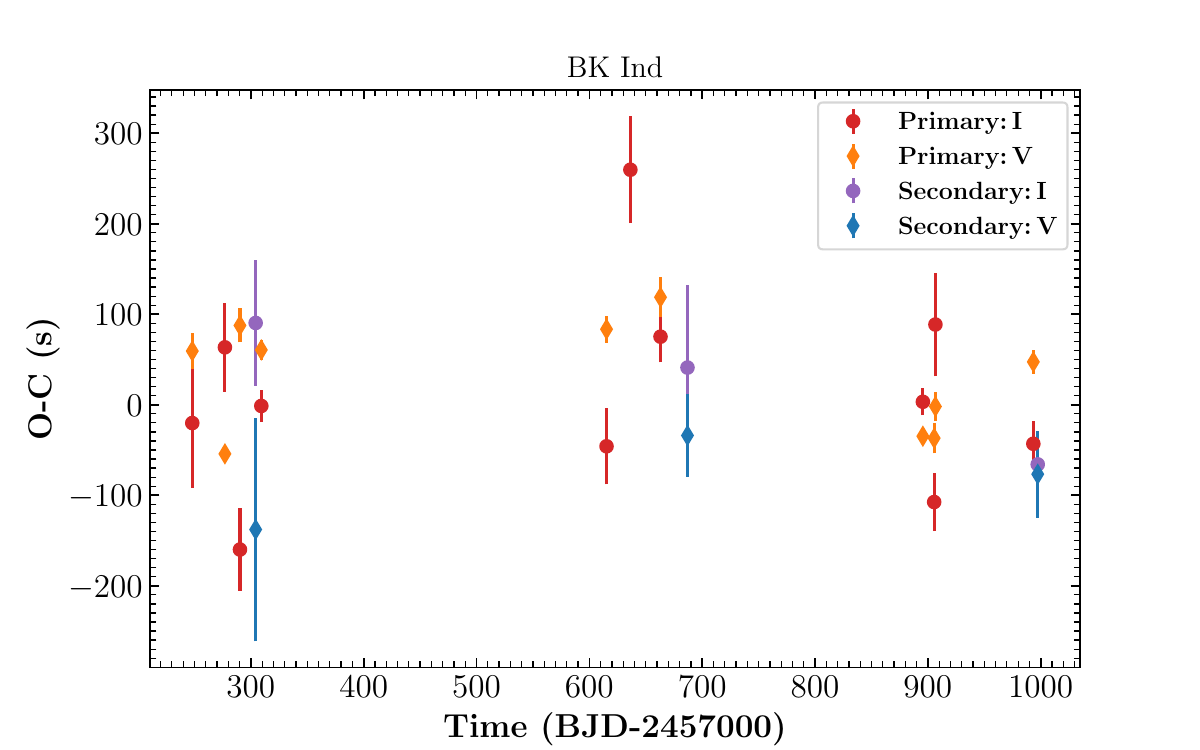} &   \includegraphics[width=0.45\textwidth]{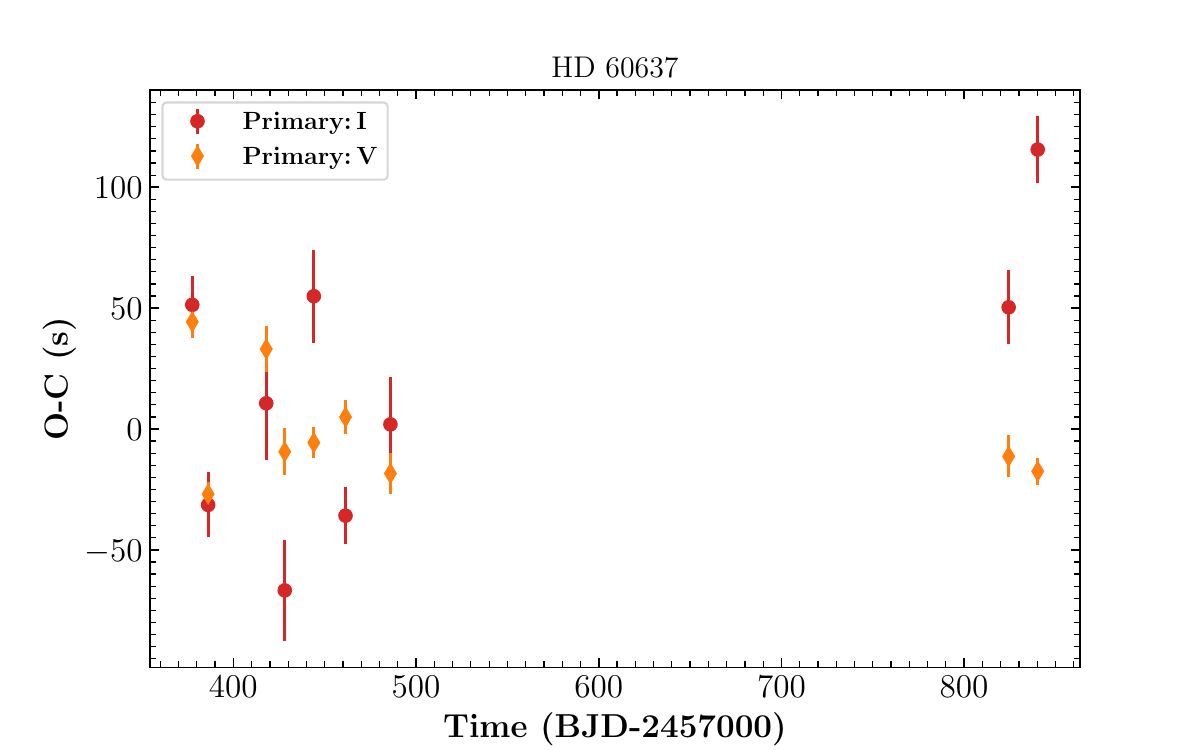} \\
(c)  BK Ind:  $P = 1.11248566$ d & (d) HD 60637:   $P = 1.4462523$ d \\[6pt]
 \includegraphics[width=0.45\textwidth]{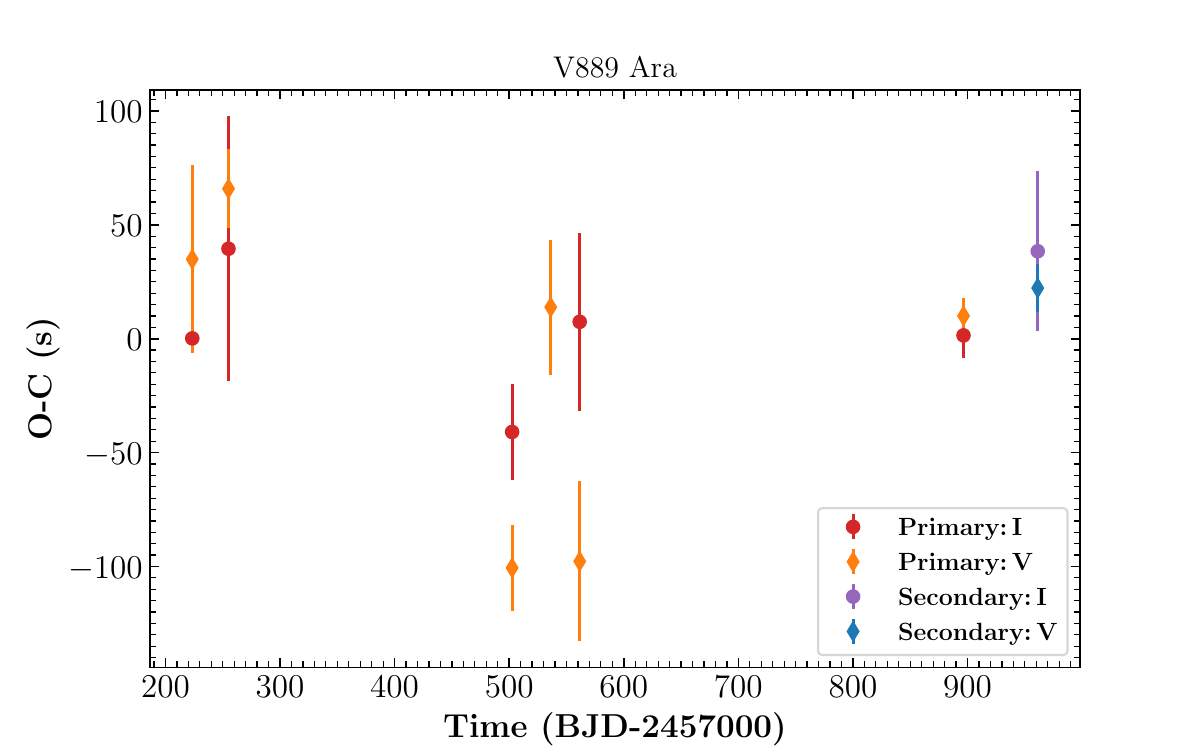} &   \includegraphics[width=0.45\textwidth]{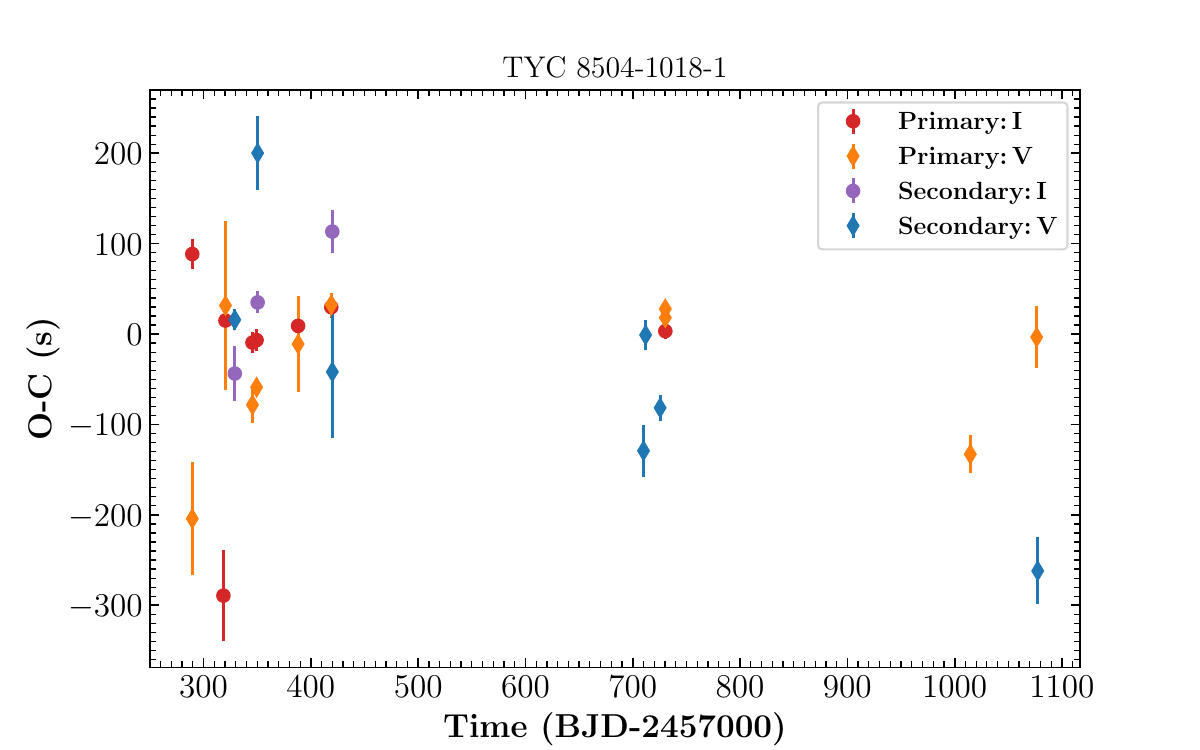} \\
(e)  V889 Ara:   $P = 1.0533183$ d & (f)  TYC 8504-1018-1:  $P = 1.9335290$ d\\[6pt]
   \multicolumn{2}{c}{\includegraphics[width=0.45\textwidth]{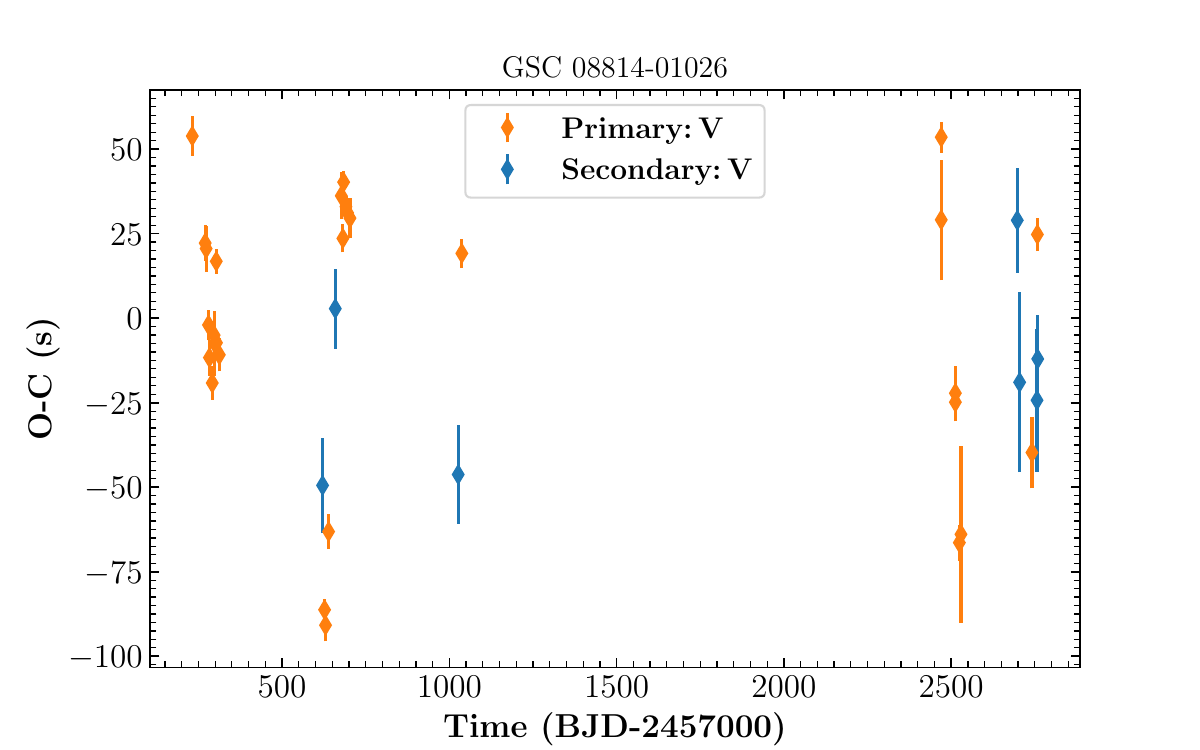}} \\
 \multicolumn{2}{c}{(g)  GSC 08814-01026 :   $P = 0.70243251$ d}\\[6pt]
\end{tabular}
\caption{\solaris{} ET for the primary (red and orange points, I and V-band respectively) and secondary eclipses (blue and purple points, I and V-band respectively) of all 7 targets.}
\label{fig:etvsol}
\end{figure*}

For the ETV plots, we first calculated reference parameters from a linear fit to cycle-number vs. $T^p_0$. This gave us reference primary epoch ($T^\mathrm{Cp}_\mathrm{0}(0)$) and estimate of the reference period ($P_\mathrm{C}$), as listed in Table.\ref{tab:etvparams}. The corresponding secondary epoch ($T^\mathrm{Cs}_\mathrm{0}(0)$) was calculated from the relations \citep{ebmilonekalrath},
\begin{equation}
    T^\mathrm{Cs}_\mathrm{0}-T^\mathrm{Cp}_\mathrm{0}-\frac{P_\mathrm{C}}{2} =\frac{P_\mathrm{C}}{\pi}e\cos{\omega}(1+\mathrm{cosec}^2{i})
\end{equation}
where  $e\cos{\omega}$ and $i$ were taken from the LC solutions.  The final ET value ($OC^j_\mathrm{X}$) for different cycle-number X was calculated as:
\begin{equation}
\label{eq:oclin}
 \begin{split}
    OC^j_\mathrm{X} &=T^j_0(X)-(T^\mathrm{Cj}_\mathrm{0}(0)+X \cdot P_\mathrm{C}) \\
    \implies OC^j_\mathrm{X} &=T^j_0(X)-T^\mathrm{Cj}_\mathrm{0}(X)
 \end{split}
\end{equation}
This gave us our final ETV plots for both primary and secondary eclipses (Fig.\ref{fig:etvsol}). 

\begin{figure*}
    \includegraphics[width=0.97\textwidth]{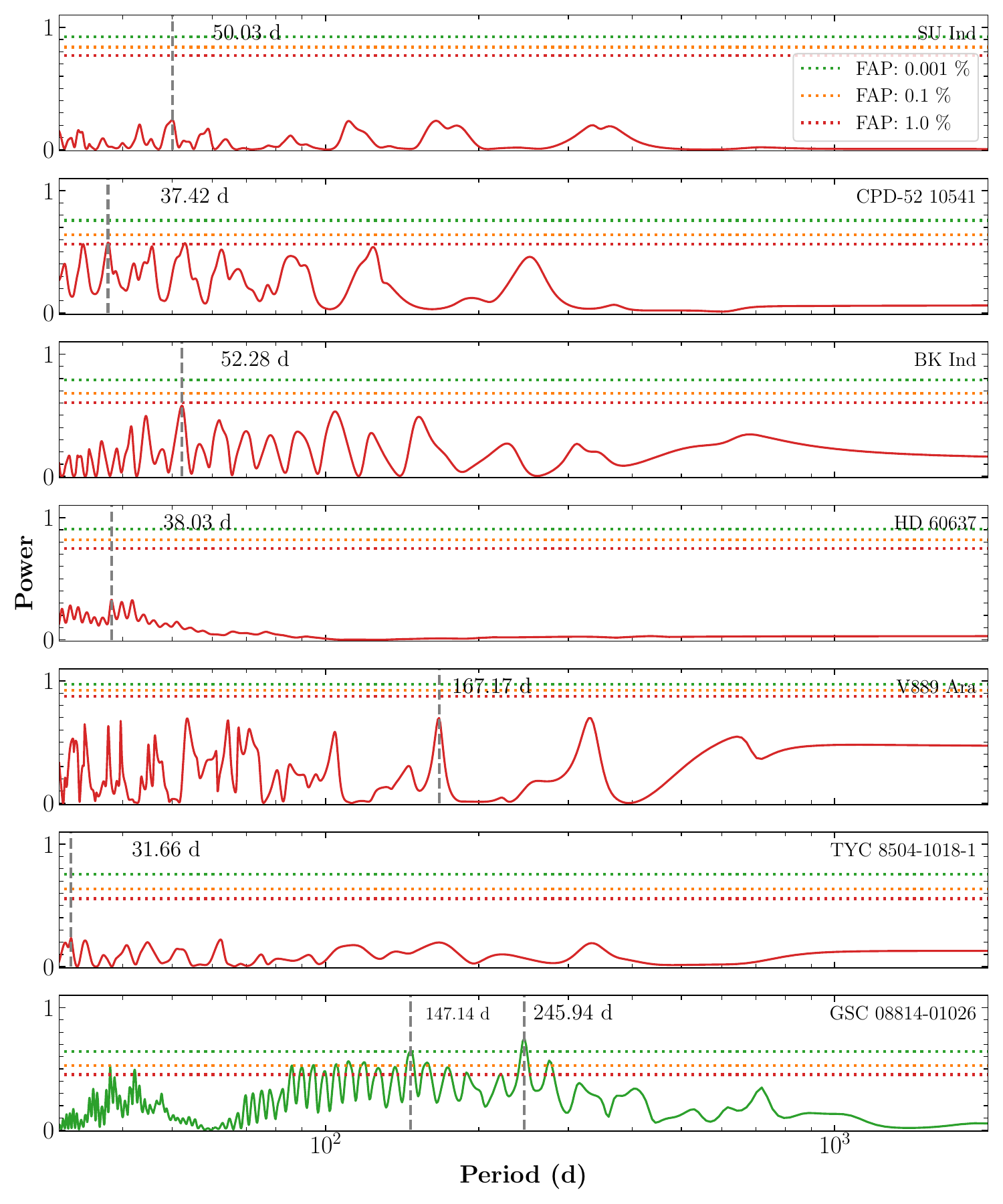}
    \caption{Lomb-Scargle periodograms for all the targets obtained using \solaris{} ET. The green, orange, and red lines represent levels corresponding to FAP less than 0.001\%, 0.1\%, and 1\% respectively. The dashed vertical lines mark the highest period visible in each of the periodograms. The green periodogram of \gsc{} marks the detection of two prominent peaks over the green level. }
    \label{fig:allperiodo}
\end{figure*}

We then took all the observed ET points and looked for periodic variations using Lomb-Scargle periodogram\footnote{https://docs.astropy.org/en/stable/timeseries/lombscargle.html} \citep{lomb,scargle}. We normalise the peaks using the standard normalisation in the periodogram to be able to compare it with false alarm levels (FAL) for every target.  We drew different FAL calculated from false alarm probabilities (FAP) using the bootstrap method in the module. The periodograms from all the targets are shown in Fig.\ref{fig:allperiodo}.

\section{Results}
\label{sec:results}
A signal detection was incurred if  $\mathrm{FAP}<0.001\%$. For $\mathrm{FAP}>1\%$, we incur no detection. 
\begin{table*}
\caption{Orbital parameters from LC fitting with \textsc{jktebop}.}
\label{tab:orbitparams}
\centering
\begin{tabular}{cccccc}
 \hline
 \hline
 Targets &  $P$  & $T_{0}$ (-2457000 BJD) & $i$  & $e\cos{\omega}$ & $e\sin{\omega}$\\
  \hline
    SU Ind & $0.9863572_{-0.0000003}^{+0.0000003}$ & $211.1200375_{-0.0000970}^{+0.0000970}$ & $81.226_{-0.353}^{+0.461}$ &$-0.00513_{-0.00027}^{+0.00028}$ & $0.01942_{-0.00390}^{+0.00455}$ \\
  \\
    CPD-52 10541  & $1.0160554_{-0.0000002}^{+0.0000002}$ & $627.8944633_{-0.0000297}^{+0.0000307}$ & $89.086_{-0.160}^{+0.193}$ & $0.00063_{-0.00006}^{+0.00006}$ & $0.00602_{-0.00079}^{+0.00077} $ \\
  \\
    BK Ind & $1.1124839_{-0.0000007}^{+0.0000007}$ & $277.0137632_{-0.0002243}^{+0.0002209}$ & $82.830_{-0.983}^{+1.762}$ & $0.00065_{-0.00099}^{+0.00099}$ & $-0.01774_{-0.01242}^{+0.01065}$ \\
  \\
    HD 60637 & $1.4462474_{-0.0000006}^{+0.0000006}$ & $373.1600044_{-0.0000679}^{+0.0000691}$ & $89.455_{-0.834}^{+0.471}$ & $0.00050_{-0.00053}^{+0.00051}$ & $0.00848_{-0.00498}^{+0.00311}$\\
  \\
    V889 Ara & $1.0533224_{-0.0000010}^{+0.0000009}$ & $255.0194406_{-0.0004040}^{+0.0004229}$ & $89.921_{-0.644}^{+0.049}$ & $-0.00020_{-0.00123}^{+0.00125}$ & $-0.04304_{-0.00647}^{+0.00642}$ \\
  \\
  TYC 8504-1018-1  &  $1.9335349_{-0.0000005}^{+0.0000005}$  &   $289.4887875_{-0.0001061}^{+0.0001061}$ & $89.084_{-0.210}^{+0.263}$ &$-0.00016_{-0.00013}^{+0.00013}$ & $0.01073_{-0.00345}^{+0.00322}$  \\
  \\
  GSC 08814-01026 & $0.7024442_{-0.0000004}^{+0.0000004}$ & $627.1265947_{-0.0000348}^{+0.0000340}$ & $87.876_{-0.098}^{+0.101}$ & 0 (fixed) & 0 (fixed)   \\
  \hline
  \hline
 \end{tabular}
 \end{table*}
 
\begin{table*}
\caption{Stellar parameters from LC fitting with \textsc{jktebop}.}
\label{tab:stellarparams}
\centering
\begin{tabular}{ccccc}
 \hline
 \hline
 Targets &  $r_1$ & $r_2$ & $J$ & $l_3$\\
  \hline
    SU Ind & $0.2845_{-0.0095}^{+0.0083}$ & $0.2885_{-0.0101}^{+0.0106}$ & $1.0416_{-0.0037}^{+0.0032}$ & $-0.017_{-0.017}^{+0.019}$ \\
  \\
    CPD-52 10541  & $0.2245_{-0.0002}^{+0.0002}$ & $0.1898_{-0.0003}^{+0.0003}$ & $0.7332_{-0.0010}^{+0.0011}$ & $0.066_{-0.003}^{0.003}$  \\
  \\
    BK Ind & $0.2275_{-0.0314}^{+0.0098}$ & $0.1722_{-0.0188}^{+0.0328}$ & $0.7715_{-0.0175}^{+0.0200}$ & $0.0183_{-0.0656}^{+0.0586}$ \\
  \\
    HD 60637 & $0.2352_{-0.0008}^{+0.008}$ & $0.1093_{-0.0014}^{+0.0012}$ & $0.2877_{-0.0039}^{+0.0038}$  &  $-0.047_{-0.023}^{+0.16}$ \\
  \\
    V889 Ara & $0.2450_{-0.0041}^{+0.0042}$ & $0.2437_{-0.0046}^{+0.0031}$ & $0.9834_{-0.0214}^{+0.0213} $ & $-0.014_{-0.011}^{+0.011}$ \\
  \\
  TYC 8504-1018-1  & $0.1326_{-0.0007}^{+0.0007}$ & $0.1055_{-0.0006}^{+0.0005}$ &  $0.7323_{-0.0220}^{+0.0221}$ &  $0.106_{-0.009}^{+0.009}$ \\
  \\
  GSC 08814-01026 & $0.2295_{-0.0012}^{+0.0013}$ &$0.1959_{-0.0018}^{+0.0023}$ & $0.595_{-0.022}^{+0.024}$ & $-0.025_{-0.013}^{+0.014}$ \\
   
  \hline
  \hline
 \end{tabular}
 \end{table*}

\begin{table*}
\caption{Linear ephemeris from fits to timing measurements.}
\label{tab:etvparams}
\centering

\begin{tabular}{cccc}
 \hline
 \hline
 Targets &  $P_\mathrm{C}$  & $T^\mathrm{Cp}_\mathrm{0}$ (-2457000 BJD) & $rms$ (s)\\
  \hline
    SU Ind & $0.9863 \pm 0.0114$& $213.1 \pm 5.5$& 91\\
  \\
    CPD-52 10541  & $1.0161\pm 0.0104$ & $241.8\pm4.8$& 64 \\ 
  \\
   BK Ind & $1.1125 \pm 0.0130$ & $248.1 \pm 5.4$& 88 \\
  \\
   HD 60637 & $1.4462 \pm 0.0167$& $377.5 \pm 2.7$& 43\\
  \\
  V889 Ara & $1.0533 \pm 0.0111$ & $223.4\pm 3.9$ &  49\\
  \\
  TYC 8504-1018-1  & $1.9335 \pm 0.0312$ &   $289.5 \pm 4.61$ &  106\\
  \\
  GSC 08814-01026 & $0.7024 \pm 0.0008$ & $231.7 \pm 1.1$ & 40  \\
  \hline
  \hline
 \end{tabular}
 \end{table*}

\begin{figure}
    \centering
    \includegraphics[width=0.45\textwidth]{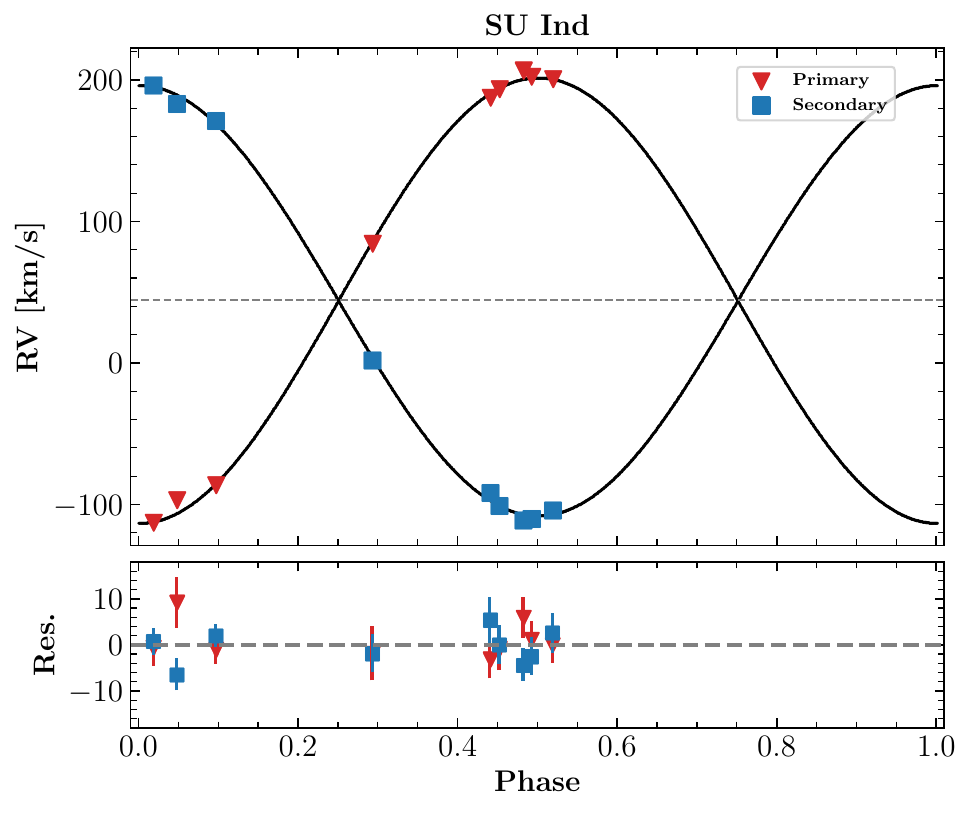}
    \includegraphics[width=0.45\textwidth]{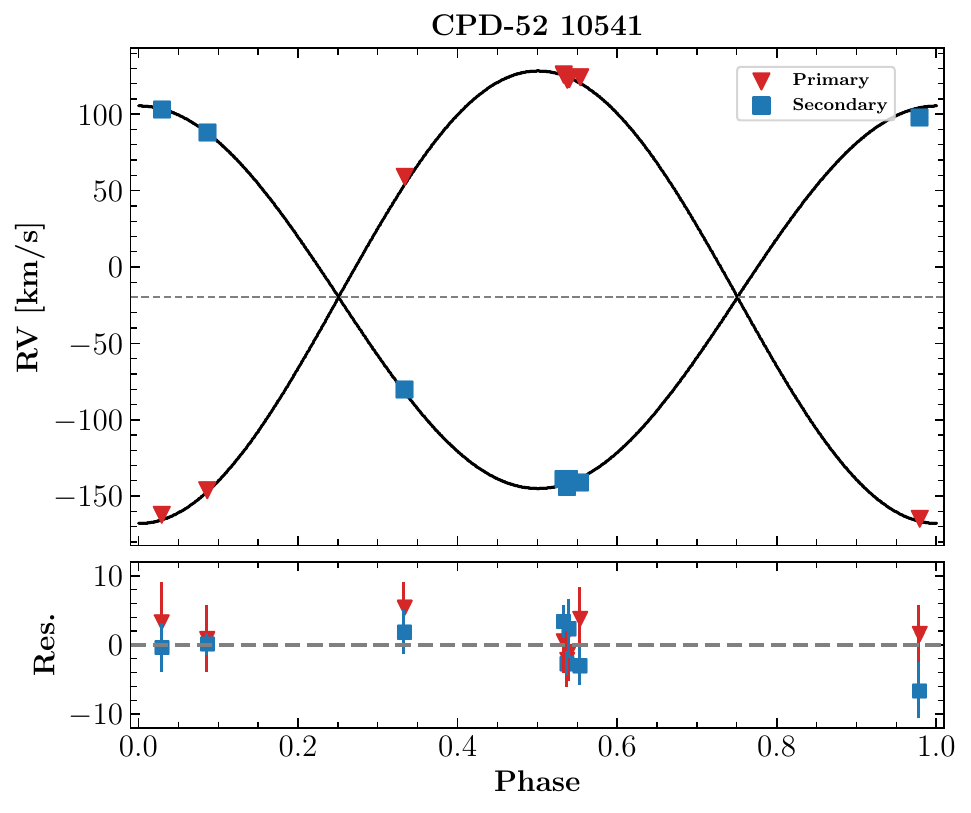}
    \includegraphics[width=0.45\textwidth]{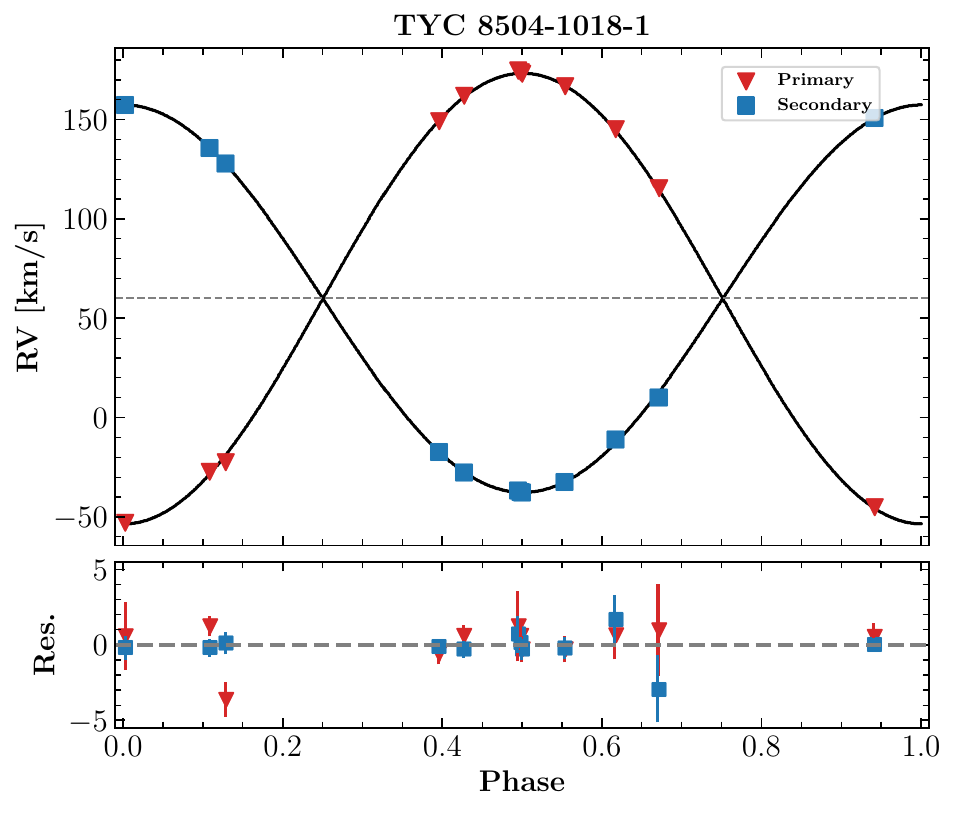}
    \caption{RV (red and blue symbols) and orbital fits (black lines) for  \su{}, \cpd{}, and \tyc{}. The corresponding residuals with their error bars are presented in the lower panels of the respective figures.}
    \label{fig:rvs}
\end{figure}

\subsection{No detection}
 
\subsubsection{SU Ind} 
 \su{} has been discovered as an eclipsing binary by \cite{hoff_su}, and is also flagged as a spectroscopic binary in the RAVE~DR6 catalogue. The only LC modelling comes from \cite{suind_budding}, who did not use any RVs, and assumed both mass ($q$) and radii ($k$) ratios to be nearly equal to 1. The derived masses and radii of both components are thus $\sim$1.20~M$_\odot$ and $\sim$1.50~R$_\odot$ (with $r_1=0.2707\pm0.0054$ and $k=r_2/r_1=0.9927\pm0.0163$). They also determined the period to be 0.9863575~d. No RV solution has been published to date, but we have acquired 9 spectra with the CHIRON spectrograph at the SMARTS 1.5-m telescope (Cerro Tololo, Chile) as a part of the Comprehensive Research with \'{E}chelles on the Most interesting Eclipsing binaries \citep[CR\'{E}ME;][]{cremepta} survey (see Appendix \ref{appx:RV}).

 Our data show that the period is 0.9863573~d. Notably, the spectroscopic mass ratio is significantly lower than 1 ($q=0.966\pm0.018$) and the (more massive) primary is fainter ($J=1.0416$), even though both components are nearly equal in size (Table~\ref{tab:stellarparams}). Our solution provides masses and radii larger than those given by \cite{suind_budding}.

 The ET measurements were possible to obtain only for the primary eclipse in the V band. Despite the large scatter of ET, no significant periodicity was found. However, the scatter is strictly dominated by measurements with larger individual errors. The $rms$ of our ET data is 91.49 sec (0.001058 d).

\subsubsection{CPD-52 10541}
\cpd{} has been discovered as an Algol-type eclipsing binary by the ASAS survey \citep[ASAS~J171606-5253.3;][]{asas0-6southern}. It has been registered in various catalogues including Gaia~DR3 but there has been no mention of its binarity.  No RV or LC analysis is available in the literature. Along with the \solaris{} photometry, we have acquired eight high-resolution spectra of \cpd{} through the CR\'{E}ME survey. The spectra were taken with the FEROS spectrograph \citep{kaufer99} attached to the MPG-2.2m telescope (La Silla, Chile), and are available in the ESO archive. We derived RV measurements and fitted the RV curves using the general methodology of the CR\'{E}ME survey (see Appendix \ref{appx:RV}).

The period of \cpd{} is 1.016055 d, and the inclination is close to $89^{\circ}$. 
The primary is larger and brighter with $J=0.7332$ and the relative radii for primary and secondary being 0.2245$\pm$0.0002 and 0.1898$\pm$0.0003, respectively. Combining the LC solution with the RV solution (Fig.\ref{fig:rvs}), gives masses and radii of $M_1=1.075\pm0.028$~M$_\odot$, $R_1=1.200\pm0.010$~R$_\odot$ and $M_2=0.908\pm0.021$~M$_\odot$, and $R_2=1.014\pm0.008$~R$_\odot$. \cpd{} has ET in both the bands (I and V) for primary and secondary eclipses. No periodicity was found in the ET. The $rms$ of the ET is 63.79 sec (0.0007~d).

\subsubsection{BK Ind}
\cite{bkindbin} was the first to classify  \bk{} as an Algol-type binary.  \cite{asas0-6southern} initially determined the period to be 1.11249 d which is close to our estimate ($P=1.1124839 d$). Recently, \cite{bkindparam} presented a full physical model, based on RV and LC, as well as a series of timing measurements collected from the literature. The derived values of masses and radii are $M_1=1.16\pm0.05$~M$_\odot$, $R_1=1.33\pm0.03$~R$_\odot$ and $M_2=0.98\pm0.04$~M$_\odot$, $R_2=1.00\pm0.03$~R$_\odot$, leading to $r_1=0.229\pm0.003$ and $r_2=0.172\pm0.004$, with inclination $i=81.97\pm0.07$~deg.

We find the inclination to be $i=82.83_{-0.98}^{+1.76}$, and the fractional radii $r_1=0.227_{-0.031}^{+0.010}$, $r_2=0.172_{-0.019}^{+0.033}$, which are consistent with the estimates in \cite{bkindparam}.  
We present new ET measurements for both primary and secondary eclipses in the I and V bands. \cite{bkindparam} concluded that there is no significant variation visible in their ET data. Similarly, we do not spot any significant period in our periodogram with \solaris{} observations. The $rms$ of our data is 88.28~sec.

\begin{figure}
    \centering
    \includegraphics[width=\columnwidth]{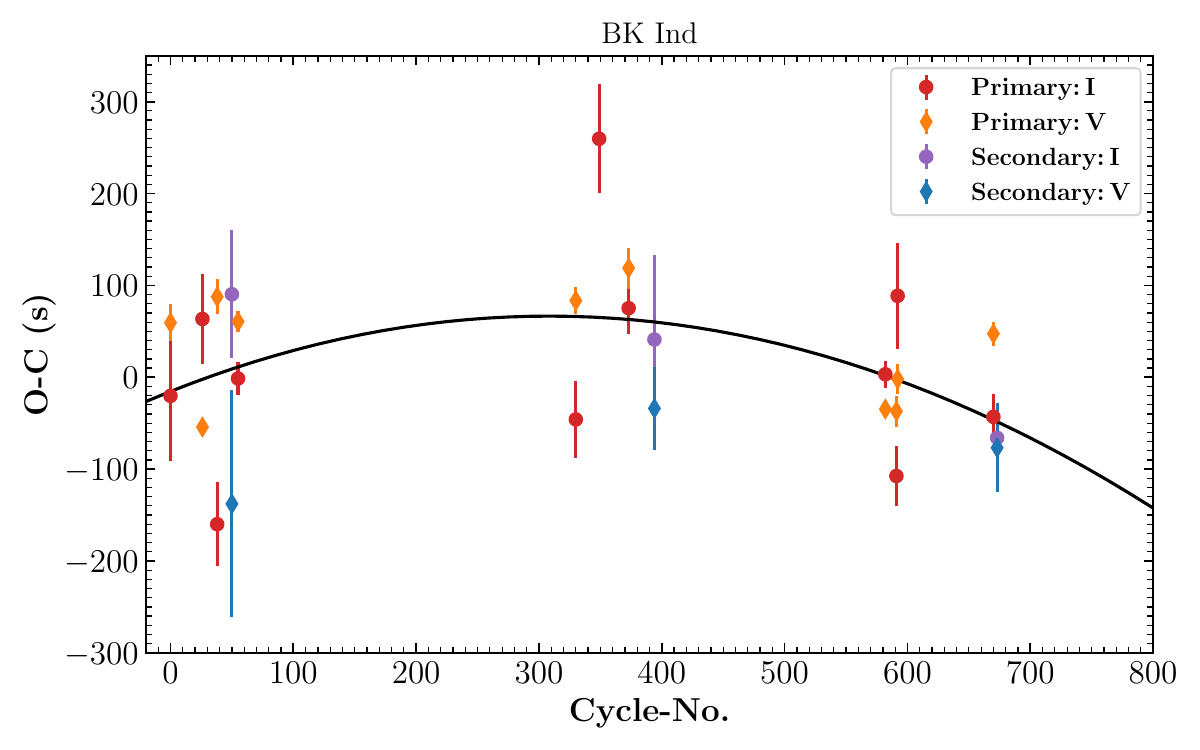}
    \includegraphics[width=\columnwidth]{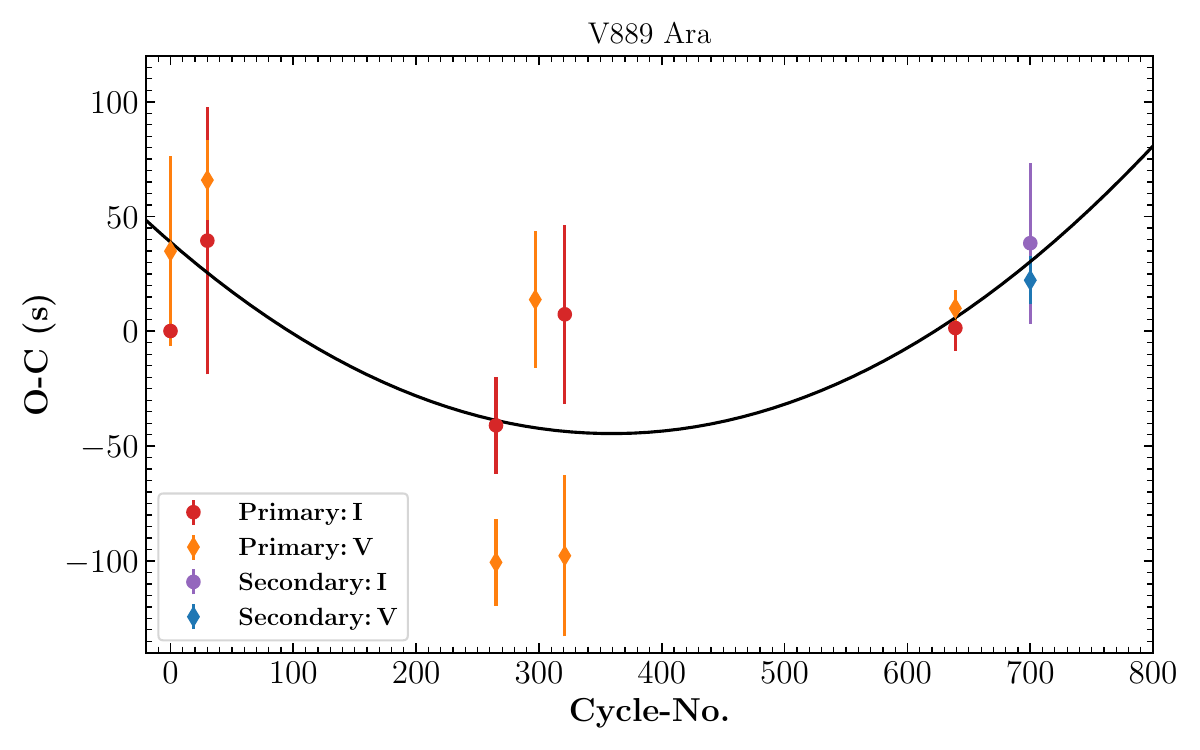}
    \includegraphics[width=\columnwidth]{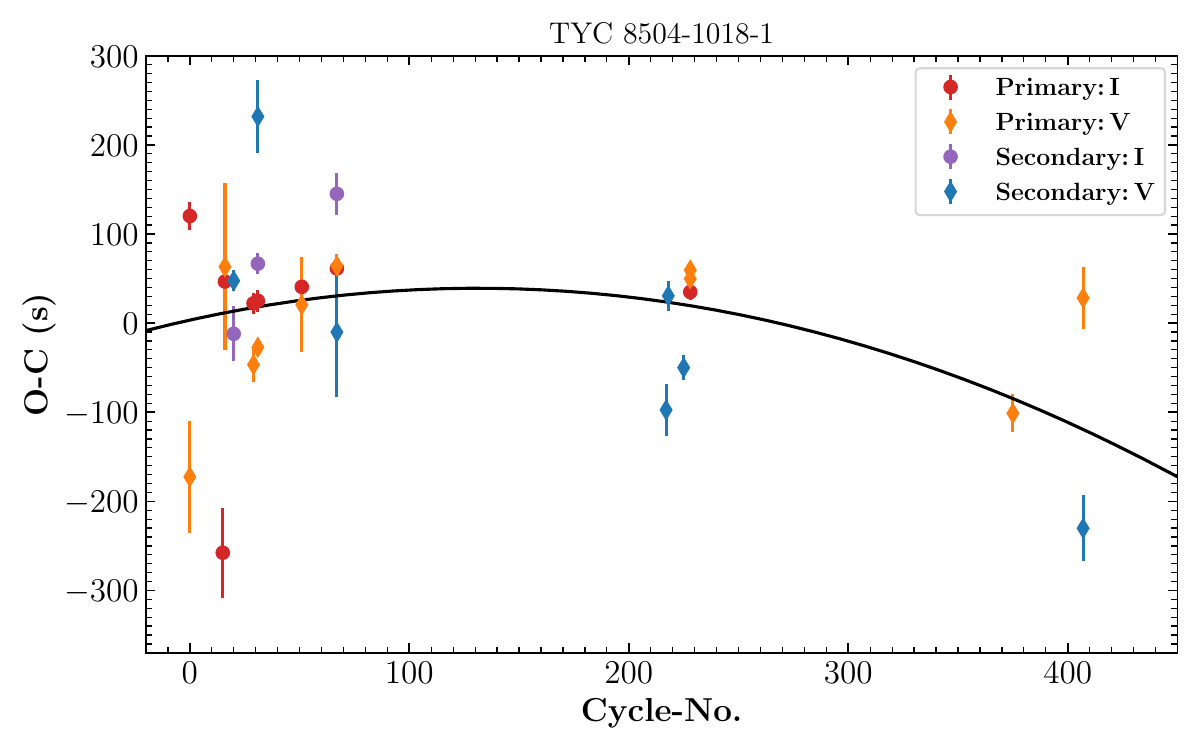}
    \caption{Parabolic fits for \bk, \vara, and \tyc{}. The respective $\dot{P}$ for the targets are -0.00172, 0.00129 and -0.0041 s/cycle.}
    \label{fig:parabolic}
\end{figure}

We found a parabolic trend in the ETV, so we fit for the rate of change of period ($\dot{P}$) as given in,
\begin{equation}
T^j_0(X)=T^\mathrm{OCj}_\mathrm{0}(0)+X \cdot P_\mathrm{OC}+\frac{1}{2}\dot{P}X^2 
\end{equation}
Substituting values from Eq.\ref{eq:oclin} we get,
\begin{equation}
 OC^j_\mathrm{X} =\frac{1}{2}\dot{P}X^2
\end{equation}
This gave us a parabolic fit to the ETV (Fig.\ref{fig:parabolic}) with the $\dot{P}=$-0.00172 s/cycle. The periodogram did not change much after removing the parabolic trend and periods were still below the detection FAL.
\subsubsection{HD 60637} 
 \hd{} has been discovered as an eclipsing binary by the ASAS survey (ASAS~J073507-0905.7), and studied in detail by \cite{hd60637_helminiak}. Their absolute masses and radii were determined based on multi-band photometry and high-resolution spectroscopy. They report the values $M_1=1.452\pm0.034$~M$_\odot$ and $R_1=1.635\pm0.012$~R$_\odot$ for the primary, $M_2=0.808\pm0.013$~M$_\odot$ and $R_2=0.819\pm0.011$~R$_\odot$ for the secondary, and an inclination of $87.91 \pm 0.01$~deg. In our analysis, we obtained fractional radii of $r_1=0.2352\pm0.0008$ and $r_2=0.1093\pm0.0013$, which is 3.5-3.8$\sigma$ distant from the results of \cite{hd60637_helminiak}. Applying their spectroscopic solution, we obtain $R_1=1.660\pm0.012$~R$_\odot$ and $R_2=0.772\pm0.011$~R$_\odot$.

The period of \hd{} is 1.4462474~d. Large differences in masses and radii between the components, and an inclination angle close to 90$^\circ$ ($89.455_{-0.834}^{+0.471}$~deg), are reflected in the fact that the secondary eclipse is shallow and total, which makes it less useful for ET measurements. Thus our \solaris{} ET are limited only to the primary eclipse in both V and I bands. No significant periodicity was identified. The $rms$ of our ET is 42.88~sec (0.0005~d). 

\subsubsection{V889 Ara}
\vara{} was registered as an eclipsing binary in the INTEGRAL-OMC catalogue of optically variable sources \citep{omcintegratl}. No LC or RV solutions have been published so far. \vara{} is a system with similar radii ($r_2/r_1=$0.9947) and surface brightness ($J=$0.9834). We have mostly primary eclipses for the system with only one secondary ET in the I-band and V-band each. No periodicity was found in the ET. With a linear fit, the $rms$ of our ET is 48.66~sec (0.0006~d). Similar to \bk{} we also see a parabolic trend here (Fig.\ref{fig:parabolic}) for which the fit gave a $\dot{P}=$0.00129 s/cycle. But even with the removal of this trend, we see no significant peak in the periodogram.

\subsubsection{TYC 8504-1018-1} 
 \tyc{} was discovered as an Algol-type eclipsing binary by the ASAS survey (ASAS~J040237-5502.5). It is flagged as a spectroscopic binary in the RAVE~DR6 catalogue \citep{rave_dr6}, but no RV or LC solutions have been published to date. It has also been observed as part of the CR\'{E}ME survey, with 8 spectra coming from the CHIRON spectrograph and 4 from the CORALIE instrument at the 1.2-m Euler telescope (La Silla, Chile). The estimated masses and radii are $1.005 \pm 0.008$ $M_{\odot}$ and $1.067 \pm 0.006$ $R_{\odot}$ for the primary. While for secondary the corresponding masses and radii are $0.864 \pm 0.004$ $M_{\odot}$ and $0.849 \pm 0.005$ $R_{\odot}$ (see the Appendix \ref{appx:RV} for details). 

The $rms$ of our ET data is 106.09~sec (0.0012~d) with the linear ephemeris. 
A fit for the slight parabolic trend (Fig.\ref{fig:parabolic}) gave us $\dot{P} =$-0.0041, but its removal did not help in finding any significant peak in the periodogram.

\subsection{Detection}

\subsubsection{GSC 08814-01026}

\begin{figure}
    \centering
    \includegraphics[width=\columnwidth]{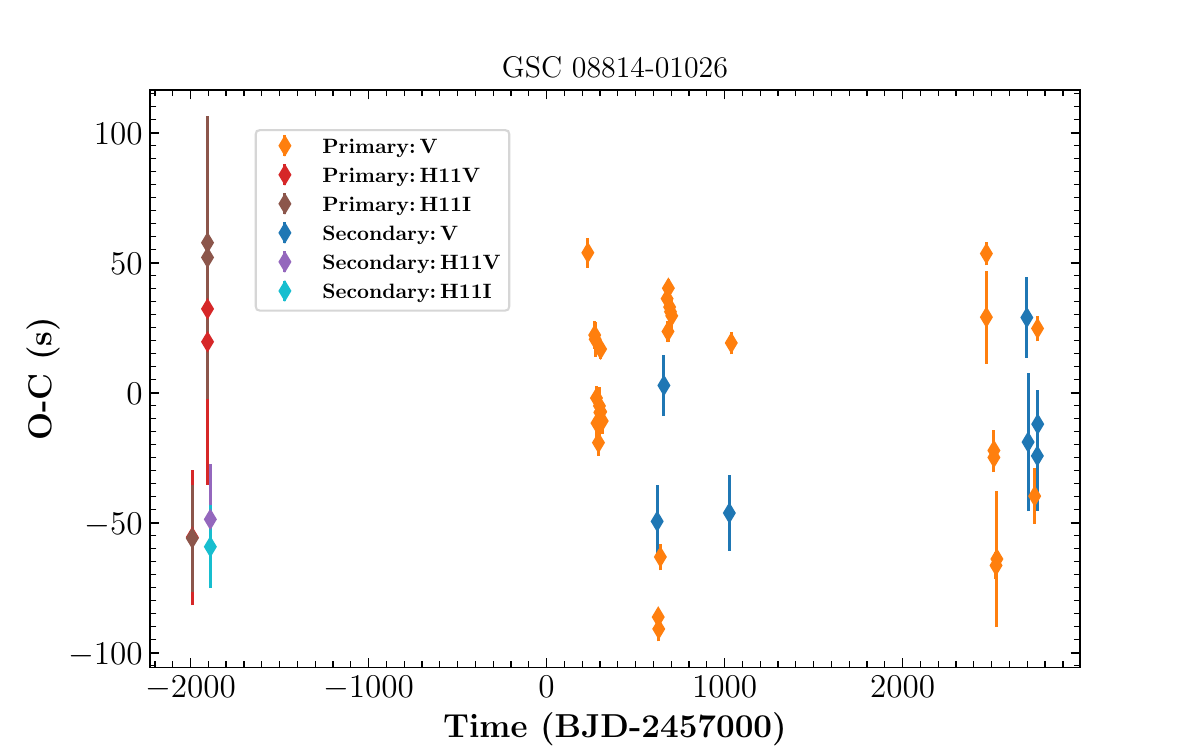}
    \caption{Additional ET for \gsc{} obtained from PROMPT and ET (marked together as OLD). The ETV were calculated using the same $T^\mathrm{OCp}_\mathrm{0}$ and $P_\mathrm{OC}$ as used in the \gsc{} ETV in Fig.\ref{fig:etvsol}.}
    \label{fig:gscallband}
\end{figure}

\begin{figure*}
    \centering
    \includegraphics[width=0.8\textwidth]{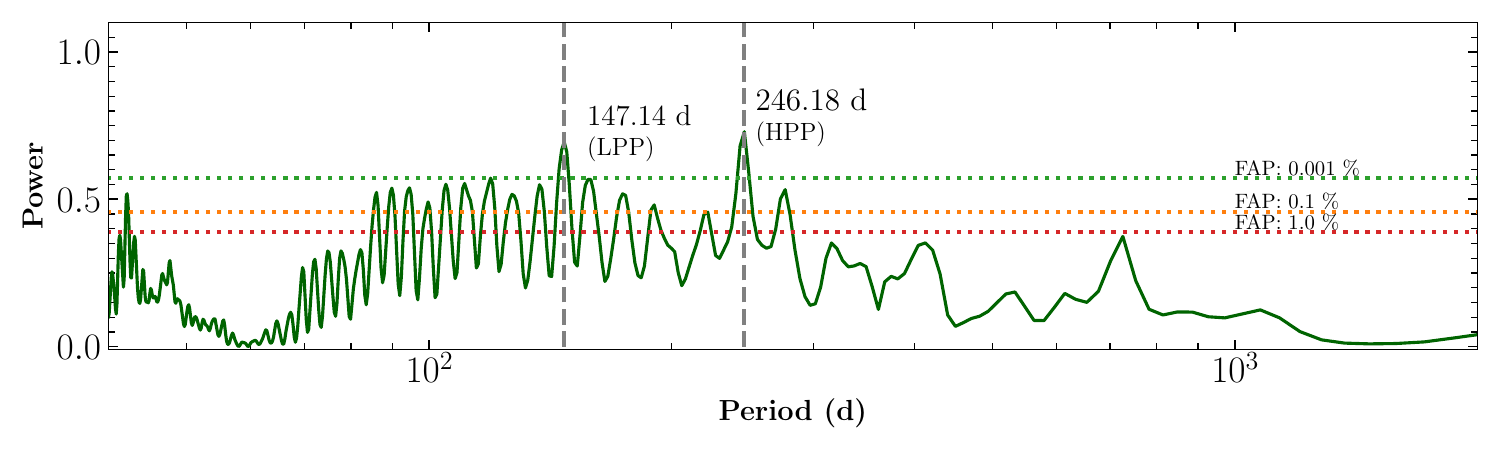}
    \caption{Lomb-Scargle periodogram for \gsc{} using all available ET points in Table.\ref{tab:gscetv}.}
    \label{fig:periodoallobs}
\end{figure*}

\begin{figure}
	\includegraphics[width=\columnwidth]{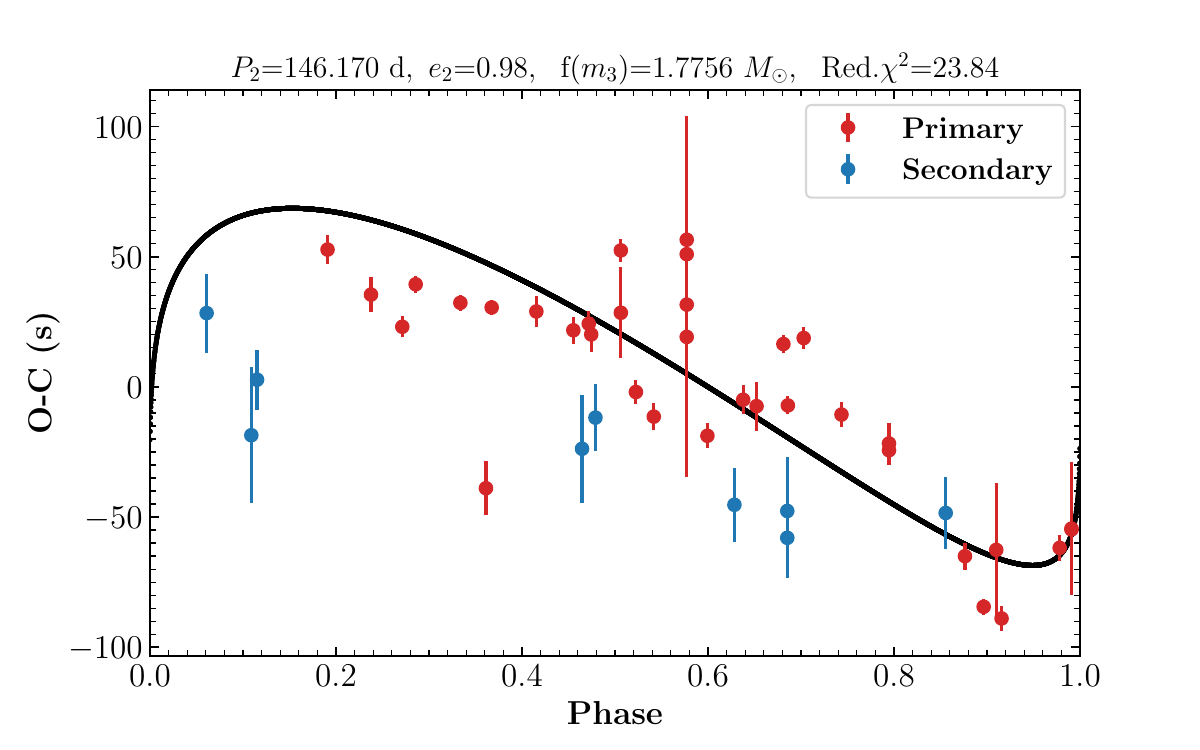}
 \includegraphics[width=\columnwidth]{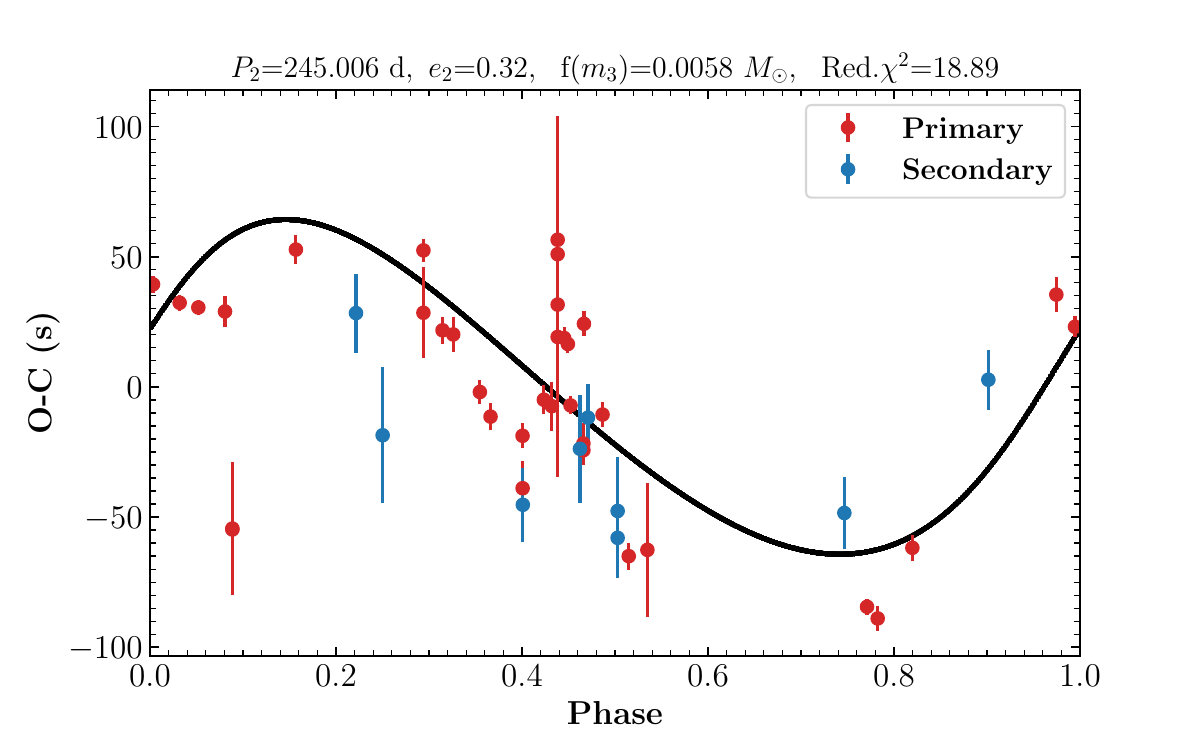}
    \caption{Phased ETV fit using \textsc{OCfit} for LPP (top) and HPP (bottom). The red points denote all the primary ET while the blue represents the secondary ET.}
    \label{fig:ocfit}
\end{figure}

%======================   OC fits  ==============================================
%=================================================================================

\begin{table}
    \caption{Parameters of orbital fits of all ET points of \gsc.}
\centering

\begin{tabular}{c|cc|cc}
\hline
\hline
\multirow{2}*{Parameter}  &        \multicolumn{2}{c}{LPP}       &       \multicolumn{2}{c}{HPP}   \\
\cline{2-5}   &     Value     & 1-$\sigma$ error  &  Value & 1-$\sigma$ error  \\
 \hline
$P_2$ (d)     &  146.170     &     0.080      &     245.006    &        0.270 \\
$a \sin{i_{2}} $ (AU)  &   0.65760    &  0.28702       & 0.13802      &      0.00373 \\
 $e_2$            &       0.979     &      0.016  &         0.320     &      0.036\\
$T_\mathrm{02}$ (BJD-2450000)     &   7498.638    &        1.436 &             5968.190    &        3.085\\
$\omega_{2}$ (deg)   &    356.148    &     1.626     &       19.196 &           3.976   \\                    
$A_3$ (s)   &      70.038    &         39.566      &        65.644    &         1.808 \\
$f(m_3)$ ($M_{\odot}$)   &  1.7756    &       2.3249    &    0.0058     &    0.0005  \\
$M_\mathrm{3,min}$ ($M_{\odot}$)   &  3.6087    &       4.3363    &    0.2668     &    0.0113  \\
$Red.\chi^{2}$ & \multicolumn{2}{c}{23.84} &\multicolumn{2}{c}{18.89} \\
\hline
\hline
 \end{tabular}
    \label{tab:ocfitgsc}
\end{table} 

%=================================================================================

This system was discovered as an Algol-type eclipsing binary by the ASAS survey (ASAS~J212954-5620.1). It has been studied in detail by \cite{helminiak_gsc}, who derived $i=87.4\pm1.3$~deg, $P_{\rm orb}=0.7024303\pm0.0000007$~d, and the following masses and radii for its components: $M_1=0.833\pm0.017$~M$_\odot$, $R_1=0.845\pm0.012$~R$_\odot$ ($r_1=0.223\pm0.002$), $M_2=0.703\pm0.013$~M$_\odot$, and $R_2=0.718\pm0.017$~R$_\odot$ ($r_2=0.186\pm0.002$). This system has the shortest orbital period and lowest stellar masses in our sample. It is also the most active one and is known to have spots that quickly evolve in time. 

The $rms$ of our ET data for linear ephemerides is 39.61~sec (0.0005~d). With just the \solaris{} observations we find few peaks with FAP $< 0.001 \%$ with a period $\sim 246 d$ dominating. This motivated us to extract minima times from three primary eclipses and one secondary eclipse from Elizabeth Telescope (ET; at South African Astronomical Observatory) and the 0.4-m Panchromatic Robotic Optical Monitoring and Polarimetry Telescopes (PROMPT; at Cerro Tololo Inter-American Observatory) in both V and I bands \citep[from ][]{helminiak_gsc}. We mark these observations as H11 in our ETV plots (Fig.\ref{fig:gscallband}). A periodogram search into this new set of ET observations makes the high-power peak (HPP) period change slightly to 246.18 d while the low-power peak (LPP) stays the same (Fig.\ref{fig:periodoallobs}). \\

To check for the possibility of a realistic orbit for these two periods, we used the O-C fitting code called \textsc{OCfit}\footnote{https://github.com/pavolgaj/OCFit} \citep{ocfit}.  For our fit, we used only the primary ET points and searched for an orbital fit based on the model of \cite{litealgorithm}. We constrained the range of sine of the semi-major axis ($a \sin{i_{2}} $) between 0.01 AU to 5 AU, and time of pericenter passage of the 3rd body ($T_\mathrm{02}$) between 2455000 BJD to 2460000 BJD. While the other parameters like eccentricity ($e_2$) and Longitude of pericenter of the orbit ($\omega_{2}$) were kept free in their physically possible ranges. To centre our searches around LPP and HPP, we restricted the tertiary period ($P_2$) around 140-150 d and 240-250 d in two separate runs. The initial optimisation was executed with Genetic algorithms \citep{geneticalgo}. We further calculated the errors using the MCMC algorithm, implemented in the code using \textsc{pymc} \citep{pymc}, which included 1000 burning iterations followed by 10,000 iterations.  This gave us two possible orbital solutions where the LPP signal has a mass-function ($f(m_3)$) 1.7756 $M_{\odot}$ and the HPP $f(m_3)$ is 0.0058 $M_{\odot}$. We estimate the tertiary mass using the equation from \cite{ocfit},
\begin{equation}
    f(m_3)= \frac{(M_3 \sin{i_2})^3}{M^2} = \frac{(a_2 \sin{i_2})^3}{P_\mathrm{2}^2}
    \label{eq:fm3}
\end{equation}
where, $M_3$ is the tertiary mass, $M$ is the total mass of the system, $a_2$ is the semi-major axis of the outer orbits and $i_2$ is the inclination of the tertiary orbit. As we do not have the estimate of $i_2$, we estimate the minimum mass of the tertiary ($M_\mathrm{3,min}$) using the total mass of the inner binary ($M_B$) from \cite{helminiak_gsc}. From Eq.\ref{eq:fm3}, we get,
\begin{equation}
    M_\mathrm{3,min}^3 - f(m_3) M_\mathrm{3,min}^2 -2 f(m_3)  M_B M_\mathrm{3,min} - f(m_3) M_B^2 =0
\end{equation}
Solving this cubic equation gives us a $M_\mathrm{3,min}$ of LPP as 3.6087 $M_{\odot}$ while that of the HPP is consistent with a M-dwarf star (0.2668 $M_{\odot}$). A $\sim$3.6 $M_{\odot}$ star would have dominated the flux in photometric (large $l_3$) and/or spectroscopic (visible in cross-correlation functions) solutions of \cite{helminiak_gsc}, which is not the case. The semi-amplitude ($A_{3}$) of the LPP signal is higher than the $A_{3}$ of HPP but its large errors bring it close to the \textit{rms}. The reduced $\chi^{2}$ of the fits favour the HPP solution. We also find that the LPP signal is not visible in the periodogram when we remove the HPP signal from the ET. The details of both the fits are tabulated in Table.\ref{tab:ocfitgsc}. \\
Since \gsc{} is an active system, these signals could likely be artefacts of star-spot migration. A study by \cite{tran_spots} showed that the effect of starspots on ETV curves can be identified by the anti-correlated nature of primary and secondary ETV curves. Therefore we over-plotted secondary ET on the phased orbital solutions for the corresponding periods to check for any visible anti-correlation (Fig.\ref{fig:ocfit}). We did not find any significant anti-correlation but since our secondary ET are sparse in number, we do not entirely rule out the possibility of the periodic signals arising due to star-spots. With the current observations though, we expect the LPP to be a result of spot migration which is corroborated by the high $e_2$ value (0.98) and larger-than-value errors on the estimated mass.

\begin{table*}
 \caption{Calculated and Observed minima times of \gsc{} used for the companion search.}
 \label{tab:gscetv}
 \centering
 \begin{tabular}{cccccc}
  \hline
  \hline
 Cycle-No.  &  Calculated $T_0$ (BJD-2457000)   &  Observed $T_0$ (BJD-2457000) &    $1 \sigma$ Error (d) & TELESCOPE & Primary/Secondary \\
 \hline
0 & 231.6578692 & 231.6584926 & 0.0000677 & \solaris :V & P \\
55 & 270.2916538 & 270.2919106 & 0.0000606 & \solaris :V & P \\
59 & 273.1013836 & 273.1016216 & 0.0000784 & \solaris :V & P \\
69 & 280.1257081 & 280.1256847 & 0.0000525 & \solaris :V & P \\
73 & 282.9354378 & 282.9353027 & 0.0000620 & \solaris :V & P \\
85 & 291.3646272 & 291.3644052 & 0.0000572 & \solaris :V & P \\
93 & 296.9840868 & 296.9840283 & 0.0000648 & \solaris :V & P \\
96 & 299.0913841 & 299.0912967 & 0.0001111 & \solaris :V & P \\
102 & 303.3059788 & 303.3061733 & 0.0000422 & \solaris :V & P \\
103 & 304.0084113 & 304.0083270 & 0.0000410 & \solaris :V & P \\
115 & 312.4376006 & 312.4374750 & 0.0000565 & \solaris :V & P \\
563 & 627.1273370 & 627.1263387 & 0.0000365 & \solaris :V & P \\
567 & 629.9370668 & 629.9360156 & 0.0000548 & \solaris :V & P \\
580 & 639.0686886 & 639.0679579 & 0.0000596 & \solaris :V & P \\
634 & 677.0000407 & 677.0004597 & 0.0000796 & \solaris :V & P \\
641 & 681.9170679 & 681.9173408 & 0.0000477 & \solaris :V & P \\
644 & 684.0243652 & 684.0248308 & 0.0000396 & \solaris :V & P \\
654 & 691.0486897 & 691.0490713 & 0.0000372 & \solaris :V & P \\
661 & 695.9657168 & 695.9660773 & 0.0000358 & \solaris :V & P \\
671 & 702.9900413 & 702.9903836 & 0.0000691 & \solaris :V & P \\
1147 & 1037.3478862 & 1037.3481078 & 0.0000505 & \solaris :V & P \\
3187 & 2470.3100785 & 2470.3106980 & 0.0000529 & \solaris :V & P \\
3247 & 2512.4560253 & 2512.4557376 & 0.0000653 & \solaris :V & P \\
3264 & 2524.3973769 & 2524.3966084 & 0.0000610 & \solaris :V & P \\
3271 & 2529.3144041 & 2529.3136641 & 0.0003034 & \solaris :V & P \\
3187 & 2470.3100785 & 2470.310415 & 0.0002064 & \solaris :V & P \\
3247 & 2512.4560253 & 2512.4557684 & 0.0000925 & \solaris :V & P \\
3573 & 2741.4490031 & 2741.4485431 & 0.0001230 & \solaris :V & P \\
3596 & 2757.6049494 & 2757.6052358 & 0.0000573 & \solaris :V & P \\
\hline
554 & 621.1569573 & 621.1563849 & 0.0001637 & \solaris :V & S \\
608 & 659.0883094 & 659.0883419 & 0.0001365 & \solaris :V & S \\
1131 & 1026.4604793 & 1026.4599441 & 0.0001679 & \solaris :V & S \\
3520 & 2704.5715957 & 2704.5713762 & 0.0003077 & \solaris :V & S \\
3510 & 2697.5472713 & 2697.5476062 & 0.0001792 & \solaris :V & S \\
3594 & 2756.5515968 & 2756.5513158 & 0.0002441 & \solaris :V & S \\
3597 & 2758.6588942 & 2758.6587546 & 0.0001521 & \solaris :V & S \\
\hline
-3163 & -1990.1359614 & -1990.1366046 & 0.0003016 & PROMPT:V & P \\
-3041 & -1904.4392028 & -1904.4388293 & 0.0003852 & ET:V & P \\
-3041 & -1904.4392028 & -1904.4389760 & 0.0006351 & PROMPT:V & P \\
\hline
-3019 & -1888.6341767 & -1888.6347400 & 0.0002448 & ET:V & S \\
\hline
-3163 & -1990.1359614 & -1990.1366088 & 0.0002379 & PROMPT:I & P \\
-3041 & -1904.4392028 & -1904.4385349 & 0.0005458 & PROMPT:I & P \\
-3041 & -1904.4392028 & -1904.4386004 & 0.0006285 & ET:I & P \\
\hline
-3019 & -1888.6341767 & -1888.6348619 & 0.0001843 & ET:I & S \\
\hline
  \hline
 \end{tabular}
 \end{table*}

\section{Conclusions}

In this work, we report the first results from the ETV search for circumbinary companions using the \solaris{} network. The accuracy and cadence are well suited for ET extraction with the possibility of getting ET with the precision of a few seconds. We extracted ET for 7 targets varying over 2 years. We report on the detection of a companion around the eclipsing binary \gsc{} about which we conclude the following: 
\begin{enumerate}
    \item There are two possible periods, 146($\pm$1) d and 245 ($\pm$1) d,  for the companion out of which the periodogram power and $\chi^2$ of orbital fit favour the 245 d period.
\item The 146 d period corresponds to an above-solar mass star (3.6087 $M_{\odot}$) while the 245 d period corresponds to an M-dwarf star (0.2668 $M_{\odot}$) orbiting around the inner binary in an orbit of eccentricity 0.32.
\item There is a possibility of the signal arising due to activity but with the current observations of the secondary ET, it seems unlikely for the 245 d period signal. However, based on the high eccentricity (0.98) of the orbital fit and the large errors in the mass estimate,  the 146 d period is more likely to be a result of stellar activity. Also, if a companion of the mass $\sim$3.6 $M_{\odot}$ existed, it would dominate the flux in the observations, which would be contradictory to \cite{helminiak_gsc}.
\end{enumerate}
With the current set of observations, we conclude that \gsc{} has an M-dwarf-like companion orbiting in an orbit of about 245 days, making it a candidate compact hierarchical triple \citep{2022borkoreview}. A follow-up with radial velocities would surely help to confirm this detection.  Nonetheless, \solaris{} proves to be a useful telescope network to look for further circumbinary companions.

\section*{Acknowledgements}

The authors thank the referee, Dr. Hans Deeg, for his comments which improved the quality of this work.
This project is funded by the National Science Center (NCN), Poland with the help of grant no.2017/27/B/ST9/02727.

AM is supported by NCN grant no.  2021/41/N/ST9/02746.  F.M. gratefully acknowledges support from the NASA TESS Guest Investigator grant 80NSSC22K0180 (PI
A. Pr\v{s}a). 

The authors would like to thank Micha{\l} Drza{\l} for his help in setting up the astrometric routine in the photometric pipeline. 

%%%%%%%%%%%%%%%%%%%%%%%%%%%%%%%%%%%%%%%%%%%%%%%%%%
\section*{Data Availability}

The minima times of all the targets are available along with the online version of this paper. The \solaris{} light curves and the radial velocities used in this work will be made available on request.



\bibliographystyle{mnras}
\bibliography{sol_phot_survey} % if your bibtex file is called example.bib

%\begin{thebibliography}{99}
%\end{thebibliography}

%%%%%%%%%%%%%%%%%%%%%%%%%%%%%%%%%%%%%%%%%%%%%%%%%%

%%%%%%%%%%%%%%%%% APPENDICES %%%%%%%%%%%%%%%%%%%%%

\appendix

\section{ET Measurements}

The ET of \gsc{} is given in Table.\ref{tab:gscetv}. The ET for the rest of the six targets are available in the online version of this paper.
%%%%%%%%%%%%%%%%%%%%%%%%%%%%%%%%%%%%%%%%%%%%%%%%%%

\section{New RV measurements and solutions.}
\label{appx:RV}
Here we describe the RVs of three systems from this study -- \su, \cpd, and \tyc{} -- which did not have any  RV solutions to date, and which were observed as part of the CR\'{E}ME survey, together with \hd{} \citep{hd60637_helminiak} and several other systems published so far.

Observations of those three systems were taken between 2012 and 2015 for \cpd{} and \tyc, and in 2021 for \su. Spectra were reduced with dedicated pipelines. RV calculations were made using our implementation of the TODCOR routine \citep{todcor}, and orbital fits with the {\sc v2fit} code \citep{konacki}. Detailed descriptions of the implemented methodology can be found in previous publications that are based on CR\'{E}ME data
\citep[see e.g.][and references therein]{krishides2017,fredv,deb11,deb12_k2,aihya,cht_ayush}.

The solutions are presented in Table~\ref{tab:creme_orb} and Figure~\ref{fig:rvs}. Individual measurements are available in the online version of this paper.

\begin{table}
    \caption{Orbital solutions of \su, \cpd, and \tyc, based on CR\'{E}ME spectra and RV measurements, and physical parameters derived from a combination of RV and LC solutions.}
    \label{tab:creme_orb}
    \centering
    \begin{tabular}{cccc}
    \hline \hline
    Parameter & \su & \cpd & \tyc \\
    \hline 
    \multicolumn{4}{l}{\it Orbital, from RVs only} \\
    $K_1$ [km/s] & 151.9(2.0) & 125.1(1.6) &  97.29(14) \\
    $K_2$ [km/s] & 157.2(2.0) & 148.0(1.4) & 113.19(43) \\
    $\gamma$ [km/s] & 44.2(1.4) & -19.5(0.9) & 60.20(19) \\
    $e$ & 0.0(fix) & 0.0(fix) & 0.0(fix) \\
    $\omega$ [$^\circ$] & --- & --- & --- \\
    $a \sin(i)$ [R$_\odot$] & 6.028(55) & 5.486(40) & 8.046(17) \\
    $M_1 \sin^3(i)$ [M$_\odot$] & 1.535(44) & 1.162(26) & 1.0044(80) \\
    $M_2 \sin^3(i)$ [M$_\odot$] & 1.483(43) & 0.982(23) & 0.8634(42) \\
    $rms_1$ [km/s] & 3.68 & 3.36 & 1.05 \\
    $rms_2$ [km/s] & 4.12 & 3.01 & 1.33 \\
    $N_{\rm obs}$ & 9 & 8 & 12 \\
    \hline
    \multicolumn{4}{l}{\it Physical, from combining with LC solutions} \\
    $M_1$ [M$_\odot$] & 1.590(46) & 1.163(26) & 1.005(8) \\
    $M_2$ [M$_\odot$] & 1.537(45) & 0.983(23) & 0.864(4) \\
      $a$ [R$_\odot$] & 6.099(56) & 5.487(40) & 8.047(17)\\
    $R_1$ [M$_\odot$] & 1.735(57) & 1.232(9)  & 1.067(6) \\
    $R_2$ [M$_\odot$] & 1.760(65) & 1.041(8)  & 0.849(5) \\
    $\log(g_1)$       & 4.161(28) & 4.323(4)  & 4.384(5) \\
    $\log(g_2)$       & 4.134(32) & 4.395(5)  & 4.517(5) \\
    \hline
    \end{tabular}
\end{table}

% Don't change these lines
\bsp	% typesetting comment
\label{lastpage}
\end{document}